\documentclass[lettersize,journal]{IEEEtran}
\usepackage{amsmath,amsfonts}
\usepackage{array}
\usepackage{textcomp}
\usepackage{graphicx}
\usepackage{booktabs}
\usepackage{multirow}
\usepackage{xcolor}
\usepackage{microtype}
\usepackage{url}
\usepackage[utf8]{inputenc}
\usepackage[T1]{fontenc}
\usepackage{cite}
\usepackage{hyperref}
\usepackage{subcaption}

\begin{document}

\title{Efficient Conversational Multi-Talker ASR with Predicted Speaker-Agnostic Activity Streams}

\author{Xiluo He, Magdalena Rybicka, Laureano Moro-Velazquez, Jes\'us Villalba,
Thomas Thebaud, and Najim Dehak%
\thanks{X. He, J. Villalba, and N. Dehak are with the Human Language Technology Center of Excellence, Johns Hopkins University, Baltimore, Maryland, USA (e-mail: xhe69@jhu.edu; ndehak3@jhu.edu).}%
\thanks{M. Rybicka is with AGH University of Krakow, Kraków, Poland. She was supported by the National Science Centre, Poland under Grants 2021/42/E/ST7/00452 and 2023/49/B/ST7/04100 and by program ``Excellence initiative -- research university'' for the AGH University of Krakow.}
\thanks{L. Moro-Velazquez is with the Department of Signal Theory and Communications, Universidad Carlos III de Madrid, Leganes, Spain.}
\thanks{T. Thebaud is with the Agence Minist\'erielle pour l'Intelligence Artificielle de D\'efense,
Palaiseau, France.}
\thanks{Work done while M. Rybicka, L. Moro-Velazquez, and T. Thebaud were with Johns Hopkins University.}
\thanks{
Xiluo He and Najim Dehak are the corresponding authors.}}

\maketitle

\begin{abstract}
    Multi-talker automatic speech recognition (ASR) can be performed by conditioning a single-speaker recognizer on frame-level speaker activity. However, this approach requires a complete diarization pass to obtain the conditioning signals and one ASR pass per speaker, making both the frontend and recognition costs dependent on the number of speakers. We reduce these costs by using Heuristic Error Assignment Training (HEAT) to merge the activities of an arbitrary number of speakers into two speaker-agnostic streams. Conditioning on these streams limits recognition to two ASR passes regardless of the number of participants. Through an analysis of the resulting encoder representations, we show that under a stream-assignment heuristic suited to conversational speech, the recognizer represents a stream target much as it does a speaker target. Because the streams do not encode persistent speaker identities, we further replace the conventional sliding-window diarization and clustering frontend with a model that predicts both activity streams directly from the audio in a single pass. A combination of arrival-order and permutation-invariant training objectives over the two-stream powerset labels prevents the predictor from collapsing to identical streams. We also investigate post-hoc speaker attribution by matching decoded utterances to speaker activities within their assigned streams. Experiments on corpora with diverse conversational dynamics demonstrate a favorable efficiency--accuracy tradeoff. On AMI and ICSI, two-stream conditioning is respectively 44\% and 87\% faster than per-speaker conditioning using the same diarization frontend, with less than 2\% relative WER degradation. Direct activity prediction increases these speedups to 79\% and 132\%, with relative WER degradations of 10\% and 21\%, respectively.
\end{abstract}


\section{Introduction}

Processing conversational speech remains one of the central challenges in automatic speech recognition (ASR). Multi-talker scenarios, such as meetings, present difficulties including overlapping speech, diverse turn-taking patterns, varying numbers of participants, and microphone conditions \cite{watanabe20b_chime, cornell23_chime, cornell2024chime, abramovski2025-chime9nsfsummary}. Large-scale pretraining and increasingly powerful sequence modeling architectures have driven remarkable progress in single-speaker recognition \cite{radford2023robust, prabhavalkar2023end, li2022recent}. Capitalizing on these improvements, target-speaker ASR has become a common paradigm for multi-talker recognition, adapting single-speaker models to conversational speech by processing the audio for each participant individually \cite{meng2024empowering, raj2021integration}. Among these, speech activity conditioned models (SAC-ASR) have emerged as a particularly effective approach, in which recognition is guided by temporal speech activity signals obtained from an external diarization or voice activity detection (VAD) model \cite{polok2024-dicowdiarizationconditionedwhispertarget, polok2024-targetspeakerasrwhisper, wang25-ssa}.

Although the target-speaker paradigm has proven effective, the scalability and practicality of these systems have received comparatively less attention as both the frontend and the ASR halves of the pipeline are expensive. On the frontend side, producing the speech activity conditioning requires full diarization: a sliding-window speaker model processes the recording in short overlapping chunks, and speaker identities are resolved through clustering. On the ASR side, the model must be executed once per participant, causing runtime to scale directly with the number of speakers. 

In this work, we address these limitations by applying Heuristic Error Assignment Training (HEAT) \cite{lu2021-surt, sklyar2021-dat, raj2023-surt2} to both sides of the SAC-ASR pipeline. HEAT merges utterances from any number of speakers into two speaker-agnostic streams, assuming at most two speakers overlap at any given time \cite{cetin06_interspeech}. On the frontend side, speaker-agnostic streams make the long-term identities that full diarization resolves irrelevant to the conditioning mask, so we predict the mask directly from the audio with a single model and remove the need for full diarization entirely. On the ASR side, conditioning on HEAT streams rather than per-speaker activities fixes the number of ASR passes at two regardless of the number of participants.

The central challenge in this framework is model collapse, which arises on both sides for different reasons. On the frontend side, the mask predictor collapses by predicting what is effectively VAD for both streams, which produces identical conditioning masks. On the ASR side, naive merging of speaker activities into streams causes both streams to decode to nearly the same transcript. The rule used in prior HEAT systems sends each utterance to the first available stream, which is usually the first, so one stream receives nearly all of the speech and the resulting imbalance leaves the conditioning too weak to steer the recognizer \cite{he2026scaling}. We examine what that conditioning writes into the encoder and find that the four mask states become mutually indistinguishable there, so the two passes are conditioned almost identically, though this does not establish why the conditioning parameters fail to differentiate them during training. We therefore introduce a training objective that anchors stream identity to arrival order, and adopt an assignment heuristic that keeps the streams balanced, to prevent collapse on the frontend and ASR sides respectively.

HEAT also discards speaker identity from the output diarization. In this manuscript, we explore whether it can be recovered after the fact by retaining the output diarization after HEAT and matching the decoded utterances against the speaker labels. Matching-based speaker attribution faces a major ambiguity at 2-speaker overlap, where both speakers span the same interval and neither can be preferred on timing alone. With HEAT, however, they have already been separated, since the assignment routes overlapping speakers to different streams, so we explore how far matching gets when each utterance is restricted to the speakers on its own stream.

This work proposes a system that addresses the practical barriers of SAC-ASR from several directions. Our contributions are as follows:
\begin{itemize}
\item \textbf{Direct prediction of HEAT conditioning masks.} We replace the sliding-window diarization and clustering frontend with a single model that predicts the conditioning masks of both streams directly from the audio. Framing the prediction as a powerset posterior yields both masks in one pass, and training with arrival-order and permutation-invariant objectives keeps the predictor from collapsing to identical streams. This substantially reduces frontend cost and makes mask sharpness a decode-time setting rather than a training choice.
\item \textbf{Analysis of HEAT-based conditioning.} We examine what stream conditioning writes into the ASR encoder, and characterize the efficiency-accuracy tradeoff across speaker counts.
\item \textbf{Post-hoc speaker attribution.} We explore labeling the output by matching utterances to diarized speakers within each stream, which separates overlapping speakers.
\end{itemize}

This paper extends our previous work presented at ICASSP 2026~\cite{he2026scaling}. We extend HEAT-based activity conditioning from the ASR model to the full pipeline, replacing the diarization frontend with direct conditioning mask prediction, examining what stream conditioning writes into the ASR encoder, and exploring post-hoc speaker attribution.

\section{Related Works} \label{sec:related}

\subsection{Multi-Talker ASR}
A common approach to end-to-end multi-talker ASR involves systems designed for a fixed number of speakers, typically targeting two-speaker scenarios. One line of work employs frontend source separation to produce speaker-specific streams before recognition, using learned masks or neural separation modules to isolate individual speakers from the mixture \cite{yoshioka2018recognizing, chen2023speech}. A related line of work proposed multi-decoder architectures that directly output transcriptions for a fixed number of speakers without explicit separation, instead relying on permutation invariant training (PIT) \cite{yu2017recognizing, yu2017permutation} objectives to assign outputs to speakers. PIT computes the training loss across all possible output-target assignments to find the optimal permutation, which becomes computationally expensive as the number of speakers grows. Both families of approaches are inherently limited in the number of speakers they can process, as the architecture must be designed for a specific speaker count. HEAT-based methods \cite{lu2021-surt, sklyar2021-dat, raj2023-surt2} avoid this constraint by fixing the number of output streams rather than the number of speakers, at the cost of discarding speaker identity. In these systems the streams are separated audio, so the assignment decides what the recognizer hears. We instead apply the same rule to speech activity, so the recognizer receives the unmodified mixture and the streams act only as conditioning, which avoids the separation artifacts that degrade downstream recognition.

Target-speaker ASR systems take a different approach, processing the recording once per participant using a single-speaker recognizer adapted to focus on a specific talker. Early approaches steered the recognizer using speaker embeddings obtained from non-overlapping speech regions or external speaker identification models \cite{saon2013speaker, kanda2019simutaneous, kanda2022-streamingembedding}. A related strategy employs frontend separation to isolate individual voices before passing them to a standard recognizer \cite{kalda2024pixit, meng2023unified}. Activity-conditioned models condition recognition on temporal speech activity from an external diarization or VAD system \cite{wang25-ssa}. In this paper, we build on DiCoW \cite{polok2024-dicowdiarizationconditionedwhispertarget, polok2024-targetspeakerasrwhisper}, which conditions a pretrained Whisper encoder on four frame-level activity states and is described in Section~\ref{sec:preliminaries}.

Serialized output training (SOT) \cite{kanda2020serialized} addressed the fixed speaker count limitation by reformulating multi-talker ASR as a single sequence generation task, interleaving the transcriptions of multiple speakers with special tokens that mark speaker changes or local speaker identities, and sequence transduction models train recognition and diarization jointly under a similar regime \cite{shafey2019joint}. The formulation has since been extended to streaming through token-level SOT \cite{kanda22_interspeech}, to large pretrained models through speaker token conditioning \cite{cornell2024joint, kocour2025adapting, wang2025_metacat}, to speech foundation models through lightweight adapters \cite{wang2024resource}, and to systems built on large language models, which leverage strong language modeling for multi-talker transcription \cite{meng2025llm, shi2025serialized, huo2026tag}. SOT removes the per-speaker pass, but is sensitive to annotation format and requires careful tuning to produce well-structured outputs across varying speaker counts \cite{subramanian2025improving}, and its extensions to target-speaker systems may still require per-speaker encoding passes to produce the activity conditioning \cite{kocour2025adapting}, partially reintroducing the cost it aims to avoid.

\subsection{Speaker Attribution for ASR}
Speaker attribution in multi-talker ASR is the task of associating transcribed content with specific speakers, at either the word or frame level. While diarization outputs can serve as an intermediate representation to guide attribution, attribution itself does not require full frame-level speaker activity estimation.

The target-speaker systems obtain frame-level speaker identity from an external diarization model or from Target Speaker Voice Activity Detection (TS-VAD) \cite{medennikov20-chime}, so attribution follows from whichever speaker the frontend recognizer was steered toward. On the other hand, joint modeling approaches integrate speaker prediction more tightly with recognition. SOT-style systems attribute implicitly, learning to associate transcribed tokens with speaker change points or local speaker indices without a dedicated prediction component, while others add an explicit speaker prediction head alongside the ASR decoder. Sortformer \cite{park2025sortformer} proposes an encoder-based diarization model that resolves the permutation problem using arrival-time sorting, ordering the per-speaker outputs by the time each speaker first becomes active so that a single pairing can be scored, and integrates speaker supervision directly into the ASR encoder via sinusoidal kernel functions. Sorting removes the need for a permutation-invariant objective, though the two are combined in practice, since arrival-time estimation becomes less reliable as the number of speakers grows. Speaker attribution has also been explored in the context of HEAT-based systems, where \cite{raj24_odyssey} investigates recovering speaker identity from the speaker-agnostic streams, demonstrating that attribution can be reintroduced into a system that discards speaker labels during training.

\begin{figure}[t]
  \centering
  \includegraphics[width=0.5\columnwidth]{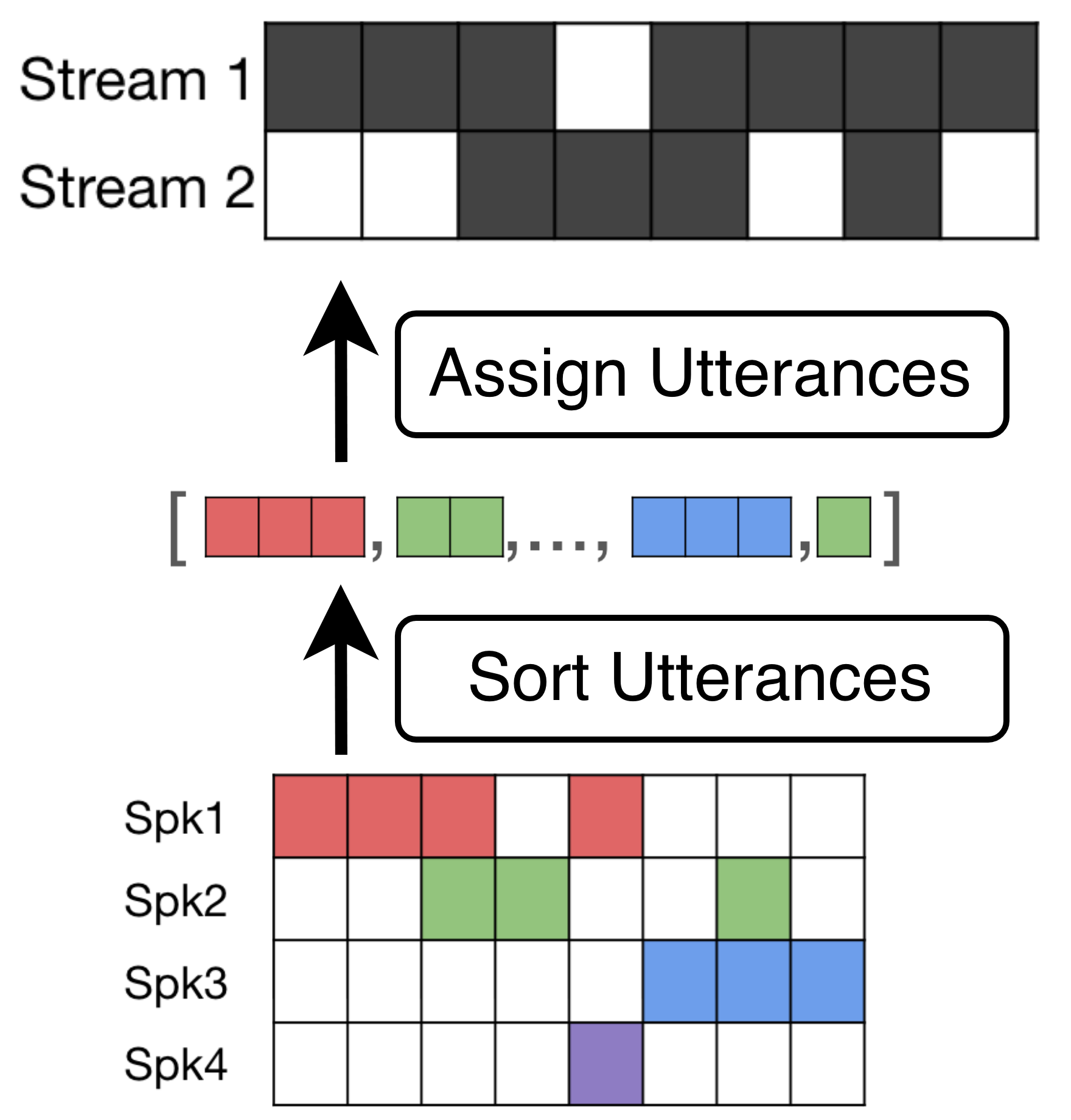}
\caption{\textbf{Heuristic Error Assignment Training (HEAT)}: Per-speaker activity assigned to one of two streams so that neither stream carries overlapping speech.}
\label{fig:heat}
\end{figure}

\section{Preliminaries} \label{sec:preliminaries}

\subsection{Heuristic Error Assignment Training (HEAT)} \label{sec:heuristic}

HEAT \cite{lu2021-surt, sklyar2021-dat} replaces the search over output-target assignments with a deterministic rule, merging the utterances of any number of speakers onto two streams under the assumption that at most two speakers are active at any given time \cite{cetin06_interspeech}.

Figure~\ref{fig:heat} shows the assignment procedure. The standard heuristic used in \cite{lu2021-surt, sklyar2021-dat} aims only to produce non-overlapping streams. A stream is available for a given utterance if it contains no other speech activity for the entire duration of that utterance, and the heuristic orders utterances by start time and assigns each to the first available stream. This guarantees non-overlapping streams, but the first stream is available for most utterances, so it receives nearly all of the speech while the second receives only what overlaps something already assigned.

Three criteria for an assignment rule suited to conversational speech are set out in \cite{he2026scaling}, which proposes the speaker-continuity heuristic to meet them. The streams should carry balanced speech activity, because single-speaker recognizers prioritize one speaker, with encoder states optimized for that speaker's accuracy \cite{wang25-ssa}, and an imbalanced pair of streams produces weight adjustments too small to counteract the bias. The standard rule produces such an imbalance, which collapses the recognizer: both passes receive near-identical conditioning and emit near-identical transcripts. The heuristic keeps the start-time order but chooses between the streams whenever both are available, so the second stream is no longer limited to overlap. The streams should also be locally continuous, which lets language model decoding draw on local context and mitigates diarization errors that fragment continuous speech \cite{linke2025context, masumura2023end}. The heuristic chooses accordingly, sending an utterance to the stream whose most recent speech came from the same speaker and otherwise falling back to whichever stream was active most recently, which preserves speaker and conversational continuity respectively.

The third criterion is that, with one or two speakers, the streams reproduce per-speaker conditioning. As originally proposed, the heuristic does so for two speakers only once they overlap, since until then the fallback places the second speaker on the first speaker's stream. We add one rule ahead of the fallback, under which each speaker's first utterance takes an empty stream where one exists. With one speaker all speech then lands on one stream, and with two speakers each owns a stream whether or not they overlap, so the HEAT mask coincides with the per-speaker mask in both cases.

\begin{figure}[t]
  \centering
  \includegraphics[width=0.9\columnwidth]{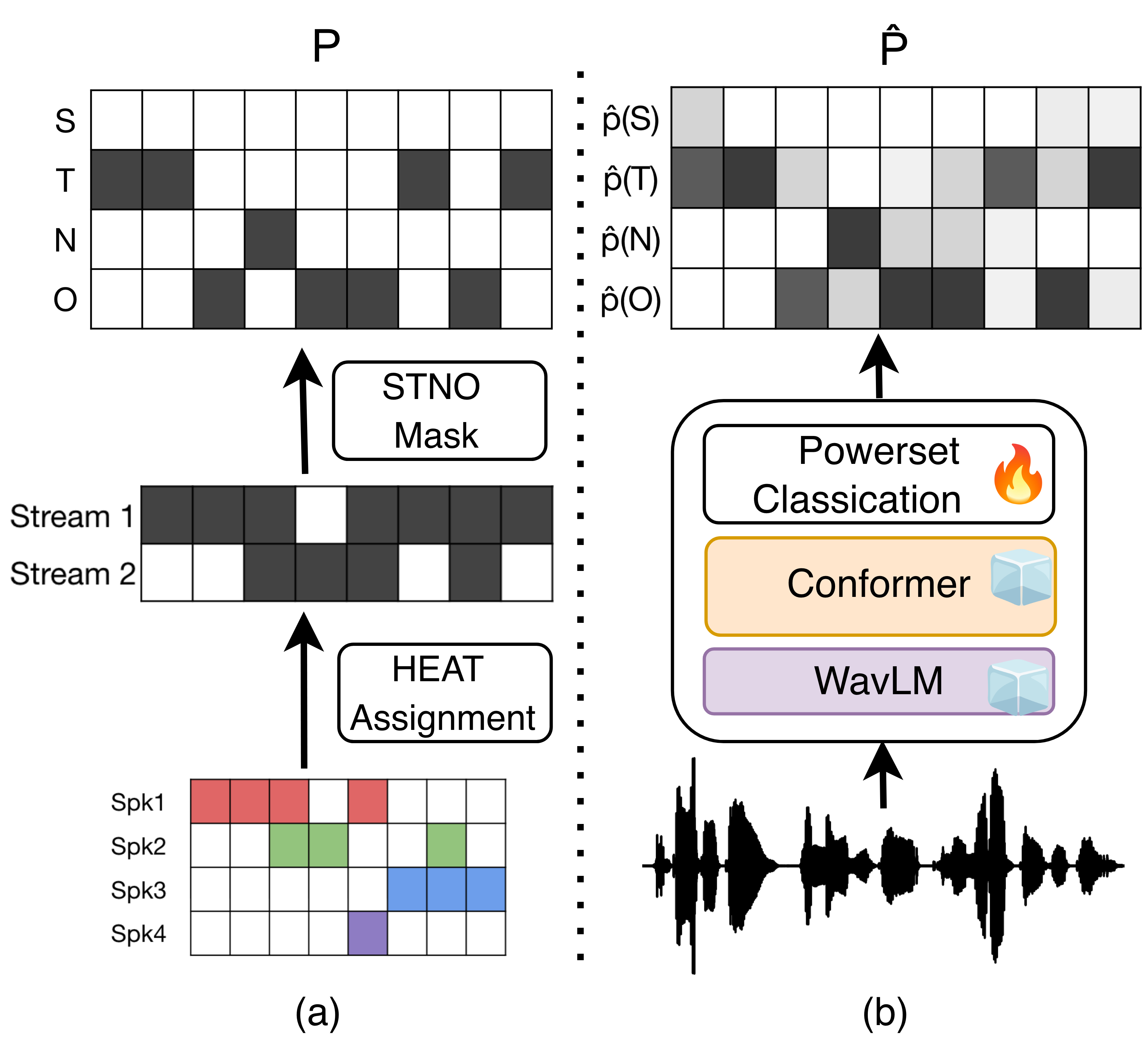}
\caption{\textbf{Training the conditioning mask predictor.} (a) The reference $\mathbf{P}$ is built from reference speaker activity: the HEAT assignment of Section~\ref{sec:heuristic} routes each utterance to one of two streams, and the STNO states follow from which streams are active at each frame. (b) The prediction $\hat{\mathbf{P}}$ comes from the audio through a frozen WavLM encoder and Conformer, followed by a trained powerset classifier.}
\label{fig:frontend}
\end{figure}

\begin{figure*}[t]
  \centering
  \includegraphics[width=0.9\textwidth]{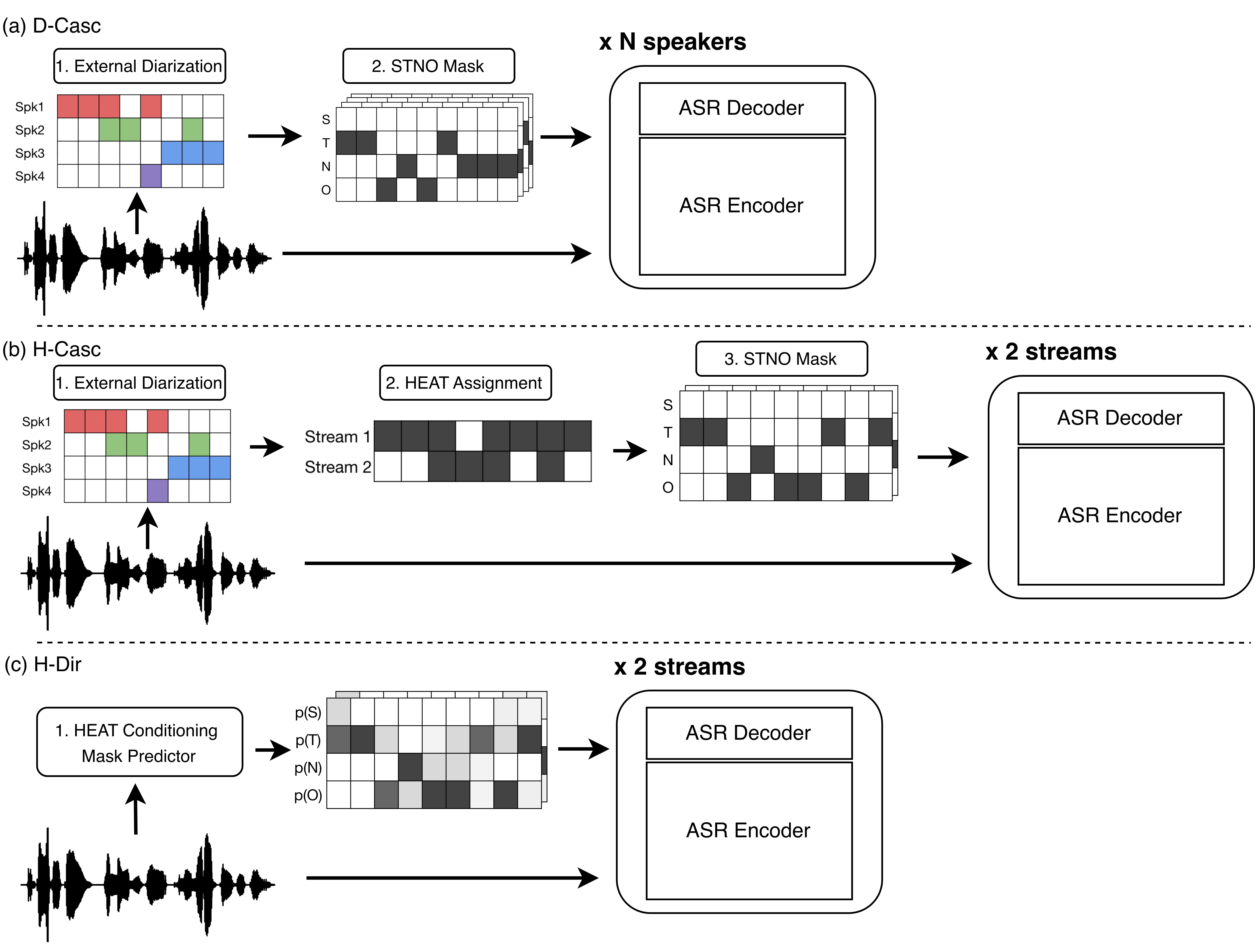}
\caption{\textbf{Deployable conditioning pipelines.} (a) D-Casc runs the recognizer once per diarized speaker. (b) H-Casc assigns the same diarization output to two streams and runs the recognizer twice. (c) H-Dir replaces diarization with a single frontend pass that predicts the masks of both streams at once, then runs the recognizer twice.}
\label{fig:asr}
\end{figure*}

\subsection{Diarization-Conditioned Whisper (DiCoW)}

DiCoW derives four mutually exclusive activity states at each frame, defined relative to a target speaker: silence, only target active, only non-target active, and overlapping speech (STNO labels). Writing $\mathcal{M} = \{\mathcal{S},\mathcal{T},\mathcal{N},\mathcal{O}\}$ for the four states, the mask at frame $t$ for a target speaker $k$ among $K$ speakers is
\begin{equation}
\mathbf{y}_{k,t} = \left[\,y_{k,t}^{(\mathcal{S})},\ y_{k,t}^{(\mathcal{T})},\ y_{k,t}^{(\mathcal{N})},\ y_{k,t}^{(\mathcal{O})}\,\right],
\qquad \sum_{m \in \mathcal{M}} y_{k,t}^{(m)} = 1 .
\label{eq:stno}
\end{equation}

Conditioning is applied via the Frame-Level Diarization Dependent Transformation (FDDT), which computes one affine transformation per state and combines them through a convex combination weighted by the mask:
\begin{equation}
f\!\left(\mathbf{h}_t, \mathbf{y}_{k,t}\right) = \sum_{m \in \mathcal{M}} \left(\mathbf{W}_m \mathbf{h}_t + \mathbf{b}_m\right) y_{k,t}^{(m)} ,
\label{eq:fddt}
\end{equation}
where $\mathbf{h}_t$ is the hidden state at frame $t$ and $\{\mathbf{W}_m, \mathbf{b}_m\}$ are learned per state. The transformation is applied at two points: once to the log-mel input before the sinusoidal positional encoding, and again before each transformer encoder layer, each application having its own parameter set $\{\mathbf{W}^l_m, \mathbf{b}^l_m\}$. We drop the layer index in our equations, since our adaptation is the same at every application point. The mask is derived from the output of an external diarization system, and the encoder is run once for each of the $K$ speakers, so the number of passes equals $K$.

\subsection{Conditioning Configurations} \label{sec:configs}

We name each configuration by its conditioning scheme and the source of its mask. The prefix D denotes per-speaker conditioning as in DiCoW, and H denotes two-stream HEAT conditioning. The suffix Orc denotes an oracle mask built from the reference annotation, Casc a cascaded mask built from the output of an external diarization pass, and Dir a mask predicted directly from the audio, as described in Section~\ref{sec:direct}. This gives the five configurations we report, D-Orc, D-Casc, H-Orc, H-Casc and H-Dir. Per-speaker conditioning has no direct variant, since its mask is indexed by speaker and needs the speaker inventory that diarization provides. The oracle configurations separate the conditioning scheme from frontend errors, and the other three are the deployable pipelines.

\section{Methods} \label{sec:method}

In this section, we describe our approach for reducing inference cost on both halves of the pipeline. The core idea is to condition the recognizer on two speaker-agnostic streams rather than on per-speaker activity, which fixes the number of recognition passes and, since the resulting mask carries no identity, makes it predictable directly from audio.

To condition an ASR model on HEAT activity, we adopt DiCoW's STNO formulation with streams in place of speakers. The four states of Eq.~\eqref{eq:stno} are defined relative to a target, so substituting a target stream for a target speaker leaves the formulation intact. Conditioning on stream $c \in \{1,2\}$ labels each frame $\mathcal{S}$ where neither stream is active, $\mathcal{T}$ where only stream $c$ is active, $\mathcal{N}$ where only the other stream is, and $\mathcal{O}$ where both are. Silence and overlap do not depend on which stream is the target, so the two streams' masks differ only by exchanging the $\mathcal{T}$ and $\mathcal{N}$ entries, an operation we write $\mathrm{swap}$. One object therefore carries both streams: we write $\mathbf{p}_{c,t}$ for the HEAT STNO mask at frame $t$ under stream $c$, $\mathbf{P}$ for a whole chunk of reference mask, and $\hat{\mathbf{P}}$ for a predicted one. The recognizer runs once per stream, so twice, regardless of how many speakers the recording contains.

We first describe how the mask is constructed, then how the recognizer consumes it, and finally how speaker labels can be recovered from the speaker-agnostic output.

\subsection{Constructing the HEAT Conditioning Mask} \label{sec:mask}

Existing HEAT systems form their streams by applying a deterministic assignment rule to reference or diarized segments. We use the heuristic of Section~\ref{sec:heuristic} throughout, and describe two ways of turning its output into a conditioning mask. The first, used by H-Casc, is a two-step construction: the assignment rule routes diarized segments onto two streams, and the STNO mask then follows from which streams are active at each frame. The second, used by H-Dir, predicts the STNO mask directly from the audio, with no segments and no assignment at inference.

\subsubsection{Construction from Diarization Output} \label{sec:mask-diar}

Let $\mathbf{x}$ denote the input audio and $\mathbf{y}_{\mathrm{spk}} = \mathrm{Diar}(\mathbf{x}) \in [0,1]^{T \times K}$ the per-speaker activity from an external diarization system. The assignment procedure of Section~\ref{sec:heuristic}, written $\mathrm{Heat}(\cdot)$, routes each speaker's contiguous stretches of activity onto two streams, giving the two-stream activity $\mathbf{y}_{\mathrm{heat}} = \mathrm{Heat}(\mathbf{y}_{\mathrm{spk}}) \in [0,1]^{T \times 2}$, which records which streams are active but not who is speaking. Reading stream $c$ as the target then fixes the four states at every frame, a deterministic map $\mathrm{Stno}(\cdot)$, so the cascaded construction is the composition
\begin{equation}
\hat{\mathbf{P}}^{\mathrm{casc}} = \mathrm{Stno}\bigl(\mathrm{Heat}(\mathrm{Diar}(\mathbf{x}))\bigr),
\label{eq:casc}
\end{equation}
the two-step construction used by H-Casc. Only $\mathrm{Diar}$ is expensive, requiring a sliding-window pass and speaker clustering, so this pipeline resolves long-term speaker identities and then discards them.

\subsubsection{Direct Prediction of the Conditioning Mask} \label{sec:direct}
Rather than performing full diarization, we train a single model $\mathrm{Mask}(\cdot)$ to predict the HEAT STNO mask directly from the audio, shown in Fig.~\ref{fig:frontend}(b),
\begin{equation}
\hat{\mathbf{P}}^{\mathrm{dir}} = \mathrm{Mask}(\mathbf{x}) \in \mathbb{R}^{4 \times T},
\label{eq:direct}
\end{equation}
which replaces the whole composition of Eq.~\eqref{eq:casc} with no diarization, no stream activity, and no assignment at inference. The assignment is still used in training, where the target $\mathbf{P}$ applies the same $\mathrm{Stno}(\mathrm{Heat}(\cdot))$ construction to reference speaker activity, as in Fig.~\ref{fig:frontend}(a). Both the cascaded and direct constructions of the STNO HEAT conditioning mask yield the same object, so they are interchangeable at the ASR conditioning step.

\paragraph{Predicting the mask as a powerset posterior}
The FDDT consumes a distribution over the four STNO states at every frame, so that is what $\mathrm{Mask}$ emits. Writing $\hat{\mathbf{p}}_t$ for the column of $\hat{\mathbf{P}}$ at frame $t$,
\begin{equation}
\hat{\mathbf{p}}_t = \left(\hat{p}^{(\mathcal{S})}_t,\ \hat{p}^{(\mathcal{T})}_t,\ \hat{p}^{(\mathcal{N})}_t,\ \hat{p}^{(\mathcal{O})}_t\right),
\qquad \sum_{m \in \mathcal{M}} \hat{p}^{(m)}_t = 1 .
\label{eq:posterior}
\end{equation}
A 30-second input at a 20\,ms frame rate gives $T = 1500$, so $\hat{\mathbf{P}}$ spans exactly one ASR input and needs no window aggregation.

For two streams this is the powerset encoding used by diarization models \cite{plaquet23_interspeech}: its four classes are the subsets that can be simultaneously active, namely neither stream, stream~1 alone, stream~2 alone, and both. Reading stream~1 as the target maps these onto $\mathcal{S}$, $\mathcal{T}$, $\mathcal{N}$ and $\mathcal{O}$.

The posterior needs sharpening before it conditions the recognizer. The predictor is trained for segmentation, so it spreads probability across states even at frames where the evidence is clear, and the FDDT weights each state's transform by its probability, so that spread weakens the target transform of Eq.~\eqref{eq:fddt}. Conditioning enters the FDDT linearly, so any transformation applied beforehand is a decode-time control requiring no retraining, and we sharpen by temperature scaling in log space,
\begin{equation}
\tilde{p}^{(m)}_t = \frac{\bigl(\hat{p}^{(m)}_t\bigr)^{\beta}}{\sum_{m' \in \mathcal{M}} \bigl(\hat{p}^{(m')}_t\bigr)^{\beta}} = \operatorname{softmax}\!\left(\beta \log \hat{\mathbf{p}}_t\right)^{(m)},
\label{eq:commitment}
\end{equation}
with a single scalar $\beta \ge 0$, which returns the raw prediction at $\beta = 1$ and the one-hot mask as $\beta \to \infty$. Because it acts on odds ratios, an intermediate $\beta$ drives confident frames to effectively one-hot while leaving near-ties graded.

The two stream masks then follow from the sharpened posterior,
\begin{equation}
\mathbf{p}_{1,t} = \tilde{\mathbf{p}}_t, \qquad \mathbf{p}_{2,t} = \mathrm{swap}\bigl(\tilde{\mathbf{p}}_t\bigr).
\label{eq:frontend-mask}
\end{equation}

\paragraph{Training objectives}
We start from the loss used to train powerset diarization models \cite{plaquet23_interspeech}. Permutation-invariant cross-entropy (PIL-CE) computes the frame-averaged cross-entropy once for every way of pairing predicted streams with reference streams, and takes the smallest of those values as the loss. The winning pairing is chosen per training example, so the model may label the streams in either order and is never charged for whichever order happens to fit. With two streams there are two pairings, differing by the exchange of Eq.~\eqref{eq:frontend-mask},
\begin{equation}
\mathcal{L}_{\mathrm{PIL}} = \min\Bigl\{\, \mathrm{CE}\bigl(\mathbf{P}, \hat{\mathbf{P}}\bigr),\;\; \mathrm{CE}\bigl(\mathrm{swap}(\mathbf{P}), \hat{\mathbf{P}}\bigr) \Bigr\},
\label{eq:pilce}
\end{equation}
where $\mathbf{P}$ and $\hat{\mathbf{P}}$ are the reference and predicted masks and $\mathrm{CE}$ the frame-averaged cross-entropy.

Taking the minimum never penalizes a labeling the model had no way to infer, which is appropriate when speaker labels are arbitrary, but it is also what allows the predictor to collapse. Distinguishing $\mathcal{T}$ from $\mathcal{N}$ is not an acoustic decision: which stream a speaker occupies is set by the assignment history, not by the signal at that frame. A model minimizing PIL-CE alone can therefore reduce its loss without ever committing to a stream, and two degenerate solutions result. Predicting only $\mathcal{S}$ and $\mathcal{O}$ reduces the mask to VAD, and predicting only $\mathcal{S}$ and $\mathcal{T}$ leaves one stream empty. Both give the two encoder passes identical conditioning, so the recognizer emits the same transcript twice.

We therefore add a second objective that resolves the permutation by convention rather than by search. Sortformer \cite{park2025sortformer} introduced arrival-time sorting for diarization, ordering reference and predicted speakers by the time each first becomes active so that one pairing can be scored and no minimization is needed, applied to per-speaker activities predicted independently of one another. Our mask is a powerset distribution over four mutually exclusive states, so we apply the same sorting to it, giving Sort-CE, and extend it with a term charging the expected stream error,
\begin{equation}
\mathcal{L}_{\mathrm{Sort}} = \frac{1}{T}\sum_{t=1}^{T} \sum_{m \in \mathcal{M}} \Bigl( -p^{(m)}_t \log \hat{p}^{(m)}_t + \lambda\, \kappa\bigl[y_t, m\bigr]\, \hat{p}^{(m)}_t \Bigr),
\label{eq:sortloss}
\end{equation}
where both sides have first been placed in arrival order, $y_t$ is the reference state at frame $t$, and $\kappa[y, m]$ is the cost of emitting $m$ when the reference is $y$. The second term charges only $\mathcal{T} \leftrightarrow \mathcal{N}$ confusions, so it is the probability mass the predictor places on the wrong stream and leaves confusions involving silence and overlap uncosted. Being linear in $\hat{\mathbf{p}}_t$ and bounded, unlike the cross-entropy beside it, at $\lambda = 0.1$ it applies steady pressure on the stream decision without dominating the objective. It is meaningful only under a fixed pairing, since minimizing over both pairings as PIL-CE does would turn every $\mathcal{T} \leftrightarrow \mathcal{N}$ confusion into a match and zero it out.

The heuristic of Section~\ref{sec:heuristic} seeds an empty stream on first use, so its targets already arrive in this order and Eq.~\eqref{eq:sortloss} anchors stream identity to the convention the heuristic follows rather than imposing a new one. Following Sortformer, we keep the permutation-invariant objective alongside the sorted one and combine the two equally,
\begin{equation}
\mathcal{L} = \tfrac{1}{2}\,\mathcal{L}_{\mathrm{Sort}} + \tfrac{1}{2}\,\mathcal{L}_{\mathrm{PIL}} ,
\label{eq:total}
\end{equation}
letting PIL-CE carry early training, before stream identity is established, while the sorted objective supplies the convention.

\subsection{Conditioning Whisper on HEAT Streams} \label{sec:heat-asr}

Substituting the HEAT STNO mask $\mathbf{p}_{c,t}$ for the per-speaker mask, Eq.~\eqref{eq:fddt} becomes
\begin{equation}
f\!\left(\mathbf{h}_t, \mathbf{p}_{c,t}\right) = \sum_{m \in \mathcal{M}} \left(\mathbf{W}_m \mathbf{h}_t + \mathbf{b}_m\right) p_{c,t}^{(m)} ,
\label{eq:heat-fddt}
\end{equation}
with the affine maps unchanged. The encoder runs once per stream, so the number of passes is two rather than $K$. Figure~\ref{fig:asr} shows the three deployable pipelines end to end.

\subsection{Recovering Speaker Attribution} \label{sec:attribution}

HEAT discards speaker identity, but the pipeline has not. Under H-Casc, the diarization pass that built the mask is still available at inference. Under H-Dir, the mask predictor keeps the frozen WavLM encoder and Conformer of the diarization model it was built from, and replaces only that model's diarization head with the mask head (Section~\ref{implementation-details}). Placing the original head back beside the mask head therefore recovers chunk-local speaker activity in the same forward pass at inference, with no additional training. We explore whether that information is enough to label the speaker-agnostic output after the fact, without a per-speaker decoding pass.

We label the output with the orchestration step of \cite{wang2024_diarizationlm}, which matches recognized speech to diarized speakers by timestamps alone. It associates each word with the speaker whose segments have the largest temporal overlap with it or, if no segment overlaps it, with the speaker at the smallest temporal distance. We apply the same rule to whole utterances, using the start and end times Whisper emits for each \cite{radford2023robust}, which makes the result less sensitive to diarization boundary errors than matching word by word. We also add an intersection-over-union term to the overlap, computed within a short margin of 0.5 seconds, so that a speaker segment that matches the utterance outranks one that merely contains it.

Timing alone cannot separate two speakers who are both active across an utterance, since both overlap it fully. HEAT has already placed such speakers on different streams, so stream-aware orchestration runs the same matching once per stream, between the utterances decoded from that stream and the speaker segments assigned to it. The co-active speaker is then on the other stream and is never a candidate. Under H-Orc and H-Casc, the HEAT assignment takes per-speaker activity as its input, from the reference or from the diarization pass, so every segment it places on a stream already carries a speaker label, which we keep. Under H-Dir the mask is predicted from the audio without a per-speaker input, so nothing links the diarized speakers to the streams, and we use stream-agnostic orchestration only.

Orchestration needs diarization with session-level speaker identities, which H-Casc has from its diarization pass. H-Dir needs two additions: one for local diarization and another for global clustering. The chunk-local diarization head supplies the speakers active within each 30-second chunk, which is long enough to separate them, but nothing ties those labels to the next chunk, so they drift apart over a recording. Linking them into session-level identities takes a sliding-window pass over speaker embeddings with a step of a third of the chunk, so each frame falls under three windows rather than the ten a full diarization pass uses.

\begin{table}[t]
\centering
\caption{Evaluation corpora measured from the reference annotation on a 10\,ms grid. $\bar{K}$ is the mean number of reference speakers per session, and the last column the proportion of speech time with three or more concurrent speakers, which the two-stream assumption cannot represent. SparseLibriMix is used at $\mathrm{ov}{=}0.1$ throughout.}
\label{tab:data-overview}
\small
\setlength{\tabcolsep}{4pt}
\begin{tabular}{lrrrr}
\toprule
Corpus & Sess. & Audio & $\bar{K}$ & \% $\ge$3-way \\
\midrule
AMI-SDM dev   &  18 &  9.67\,h & 4.00 &  1.4\% \\
AMI-SDM test  &  16 &  9.06\,h & 3.94 &  2.4\% \\
ICSI-SDM test &   3 &  2.77\,h & 7.00 &  0.5\% \\
\midrule
NOTSOFAR-1 dev & 117 & 12.17\,h & 4.68 &  3.6\% \\
CHiME-6 eval   &   2 &  5.21\,h & 4.00 &  8.4\% \\
\midrule
SparseLibriMix $K{=}2$ & 500 & 1.18\,h & 2.00 & 0.0\% \\
SparseLibriMix $K{=}3$ & 500 & 1.13\,h & 3.00 & 1.3\% \\
\bottomrule
\end{tabular}
\end{table}

\section{Data and Experimental Setup} \label{sec:setup}

\subsection{Training data} \label{sec:train-data}
We train the conditioning mask predictor on the joint AMI \cite{kraaij2005ami}, AliMeeting \cite{Yu2022M2MeT, Yu2022Summary}, and AISHELL-4 \cite{fu21b_interspeech} corpus following the DiariZen protocol \cite{han2025diarizen}, which covers a broader range of meeting conditions and speaker configurations than AMI alone. We train the ASR model on AMI only, in the single distant microphone (SDM) condition, which comprises approximately 100 hours of multi-speaker meetings with four to five participants per session.

\subsection{Evaluation data} \label{sec:data}

We evaluate on AMI, ICSI, NOTSOFAR-1, CHiME-6 and SparseLibriMix \cite{kraaij2005ami, janin2003icsi, cornell2024chime, watanabe20b_chime, cosentino2020librimix}, all single-channel, using the single distant microphone (SDM) condition for AMI and ICSI and reducing the CHiME-6 arrays to one distant channel. Table~\ref{tab:data-overview} reports statistics measured directly from the evaluation manifests. AMI is our primary development corpus, on which we perform model selection and tune the sharpening parameter $\beta$ using the dev split and report meeting-domain results on the test split.

The remaining corpora each isolate a property the two-stream assumption is sensitive to. ICSI has the largest speaker inventory and the least concurrent speech of any real corpus we use, which separates the number of speakers a system must distinguish from the amount of simultaneous speech it must transcribe. NOTSOFAR-1 gives 117 independent sessions with a varying number of participants, so aggregate word error carries tighter confidence and we can observe systems whose cost depends on the speaker count across a mixed population. CHiME-6 is our stress case, with sessions averaging 2.6 hours, nearly three times the length of the next longest corpus we use, and the highest rate of three-or-more-speaker overlap; we use the eval split. We note that the diarization frontend hypothesizes only 66\% of the annotated reference speech there, against 92 to 96\% elsewhere, which flatters error rates dominated by insertions and substitutions, so we read its results as a difficulty check rather than as a system comparison.

To evaluate performance under controlled speaker count and overlap, we use SparseLibriMix, which mixes clean LibriSpeech \cite{panayotov2015_librispeech} utterances on a timeline with controlled silence and a controlled degree of speaker overlap. We use $K \in \{2,3\}$ speakers at the overlap setting $\mathrm{ov}{=}0.1$, both exact by construction. At $K{=}2$ the HEAT mask coincides with the per-speaker mask (Section~\ref{sec:heuristic}), and $K{=}3$ is the smallest count at which speakers must share a stream.

\subsection{Implementation Details} \label{implementation-details}

\subsubsection{Diarization Frontend}
Both cascaded configurations obtain per-speaker activity from DiariZen \cite{han2025diarizen}, which we run at the 16\,s window and tenfold overlap it was released with, followed by its speaker clustering stage. D-Casc conditions on that output directly and H-Casc applies the heuristic of Section~\ref{sec:heuristic} to the same output, so the two differ only in the conditioning scheme. The mask predictor described next is built on the same checkpoint, so the frontend it replaces and the backbone it inherits are one model.

\subsubsection{HEAT Conditioning Mask Prediction Model}
The mask predictor follows the DiariZen architecture. The acoustic backbone is a WavLM-Large encoder (24 transformer layers plus the convolutional feature extractor, yielding 25 hidden representations of dimension 1024), and both it and the 4-layer Conformer above it are initialized from a publicly available diarization-adapted checkpoint and kept frozen, so the mask prediction head, which replaces the checkpoint's diarization head, is the only component trained. We obtain frame-level features via a learnable softmax-weighted sum across all WavLM layers, projected to 256 dimensions and layer-normalized before the Conformer, which uses attention dimension 256, feed-forward dimension 1024, 4 attention heads, and dropout 0.1. We use no positional encoding, and pass the Conformer output to a linear classifier head that produces the four-state posterior $\hat{\mathbf{P}}$ of Eq.~\eqref{eq:posterior}. At decode time we sharpen the posterior with the operator of Eq.~\eqref{eq:commitment} at $\beta{=}16$, selected by sweeping $\beta$ on the development data and applied unchanged across all corpora.

The training objective is Eq.~\eqref{eq:total}. We fine-tune with AdamW using a one-cycle schedule (10\% warmup ratio) and peak learning rate $1 \times 10^{-3}$, for 10 epochs on the joint AMI, AliMeeting, and AISHELL-4 corpus. During heuristic assignment we stochastically perturb segment boundaries by up to 10 frames with probability 0.5.

\subsubsection{ASR Model: DiCoW v3.3}
To keep the comparison with speaker conditioning fair, the recognizer, its conditioning interface, and its training recipe are those of DiCoW throughout, and the only difference between the two systems is whether the mask indexes a speaker or a stream. In line with DiCoW, we use Whisper-large-v3-turbo with the DiCoW v3.3 conditioning interface \cite{polok2024-targetspeakerasrwhisper, polok2024-dicowdiarizationconditionedwhispertarget, polok2026_sedicow}, applying FDDT layers at every encoder layer with diagonal projection and the disparagement initialization that biases non-target and silence frames toward attenuation, and a pre-positional FDDT layer conditioning the log-mel input prior to the sinusoidal positional embedding. Following DiCoW, we mount a Connectionist Temporal Classification (CTC) head on the encoder output and combine it with the standard cross-entropy decoder loss at a CTC weight of 0.3, stripping timestamp tokens from the CTC target.

We train with AdamW at a base learning rate of $5 \times 10^{-7}$ and a $100\times$ learning-rate multiplier on the FDDT parameters, using a 100-step linear warmup followed by cosine decay and an effective batch size of 64. We train for up to 10 epochs capped at 50k steps in bf16 precision, with gradient clipping at 1.0 and no weight decay. We augment both the audio and the conditioning mask with SpecAugment ($p{=}0.3$), MUSAN additive noise ($p{=}0.3$) \cite{snyder2015_musan}, Gaussian-noise perturbation of the mask (variance 0.2, $p{=}0.75$), and segment-flip augmentation ($p{=}0.3$, per-segment change probability 0.1, segment lengths 5--50 frames). Both systems are conditioned on masks built from the same segments, HEAT on streams assigned by the heuristic of Section~\ref{sec:heuristic} and DiCoW on the speakers those segments came from.

\subsubsection{Speaker Attribution}
We label each decode with the diarization matching its own mask source, recovering the stream assignment record offline from the preprocessing for H-Orc and H-Casc. The chunk-local diarization head is the original head of the DiariZen checkpoint, reattached without retraining, and predicts up to $K_{\max}{=}4$ speakers per chunk, fixed across all corpora. Under H-Dir, we lift these to session-level identities by clustering speaker embeddings from a pretrained WeSpeaker \cite{wang2023_wespeaker} ResNet-34 model over a sliding window with a 10\,s step, extracted from sub-segments of 1--2\,s on frames where the frontend predicts a single active speaker and the ASR output attests speech, with overlapping regions excluded. We use agglomerative clustering on cosine distance with soft must-link constraints within each chunk-local speaker, and tune the threshold on the AMI development split.

\subsection{Evaluation Metrics} \label{evaluation-metrics}

\paragraph{Frontend error terms}
We measure frontend quality along two axes: how much speech the system finds, and how faithfully it reproduces the assignment the heuristic would have made. We report these as separate terms rather than a single aggregate. False alarm is speech predicted where none exists and missed speech is speech present but not predicted. Both are defined identically for diarization and for mask prediction, since the union of the two HEAT streams is the total predicted speech, and both have direct recognition consequences: false alarm becomes insertions and missed speech becomes deletions.

The third component of diarization error is confusion, detected speech given the wrong label, which for diarization is the wrong speaker among $K$ candidates and for mask prediction the wrong one of two streams. Confusion is frame-weighted, which suits speaker labels but not stream labels: stream identity is set by the assignment history rather than by the signal at each frame, so a single wrong decision early in an utterance is charged for every frame that follows it, while a brief flip to the other stream and back is charged for almost nothing. We therefore also count stream swaps, the moments at which speech moves to the other stream, reported per minute. A model swap with no reference swap within 0.5\,s is unnecessary, typically a flip to the other stream and back within a single utterance. A reference swap the model never makes is missed, a stream change it does not reproduce at all. Confusion mixes the two in one number, whereas the swap counts say which of them an objective acts on.

\paragraph{Word error rates}
We report two word error rates (WERs), both computed with MeetEval~\cite{MeetEval23} in a time-constrained form that forbids aligning hypotheses and references across implausibly long temporal distances, for which we use a 5-second collar. Where the output carries no speaker labels we report tcORC-WER, the time-constrained \cite{vonneumann25_tcorcwer} form of optimal reference combination WER \cite{vonneumann2023-orcwer}, which searches over assignments of reference utterances to output streams subject to each speaker's utterances staying in order. Where the output is speaker-attributed we report tcpWER, which concatenates the utterances of each speaker in both reference and hypothesis and scores the permutation between them that minimizes WER. Read against tcORC-WER on the same output, tcpWER isolates the error attributable to speaker assignment rather than to transcription. We micro-average both over the sessions of a corpus, dividing total errors by total reference words, and report tcORC-WER unless labeled otherwise.

\paragraph{Inverse Real-Time Factor (RTFx)}
RTFx measures inference efficiency as the ratio of audio duration to total processing time, including all preprocessing, encoding, decoding, and postprocessing:
\begin{equation}
\mathrm{RTFx} = \frac{\sum_{i=1}^{N} T^{(i)}_{\mathrm{audio}}}{\sum_{i=1}^{N} T^{(i)}_{\mathrm{proc}}}.
\end{equation}
Higher RTFx indicates faster processing. The RTFx of a cascaded configuration includes its diarization pass, and that of H-Dir includes the mask predictor, which runs once per chunk and serves both streams. The oracle configurations have no frontend, so their RTFx is an upper bound rather than a deployable figure.

\begin{table}[t]
\centering
\caption{\textbf{Frontend prediction quality on AMI-SDM test.} FA, Miss and Conf.\ are the three components of diarization error, as percentages of reference speech. Conf.\ is detected speech given the wrong label: the wrong speaker among $K$ for DiariZen, the wrong one of two streams for the mask predictor. The last two columns count stream swaps per minute, unnecessary and missed respectively, and are specific to two-stream assignment. DiariZen is shown at its native 16\,s window and again at the 30\,s window the mask predictor uses.}
\label{tab:frontend}
\footnotesize
\setlength{\tabcolsep}{7.0pt}
\begin{tabular}{lrrrrr}
\toprule
& & & & \multicolumn{2}{c}{Stream swaps} \\
\cmidrule(lr){5-6}
System & FA$\downarrow$ & Miss$\downarrow$ & Conf.$\downarrow$ & Unnec.$\downarrow$ & Missed$\downarrow$ \\
\midrule
\multicolumn{6}{l}{\emph{DiariZen}} \\
16\,s window         & 3.99 & 6.04 & \phantom{0}2.23 & --- & --- \\
30\,s window         & 3.75 & 5.87 & \phantom{0}2.79 & --- & --- \\
\midrule
\multicolumn{6}{l}{\emph{HEAT-STNO, 30\,s window}} \\
PIL-CE only          & 5.20 & 7.26 & 10.91 & 3.41 & 2.93 \\
\; + Sort-CE         & 5.03 & 3.98 & \phantom{0}9.58 & 3.12 & 3.04 \\
\quad + stream error & 4.20 & 4.57 & \phantom{0}7.50 & 2.64 & 3.14 \\
\bottomrule
\end{tabular}
\end{table}

\section{Results}
\label{sec:results}

We first evaluate each half of the pipeline on its own. Section~\ref{sec:results-mask} measures the mask predictor against the diarization frontend it replaces, and Section~\ref{sec:results-conditioning} examines what stream conditioning writes into the recognizer's encoder and how the assignment rule shapes it. Section~\ref{sec:results-main} then reports the speedup and accuracy cost of each configuration, and Section~\ref{sec:results-attribution} explores post-hoc speaker attribution.

\begin{figure*}[t]
  \centering
  \begin{subfigure}{0.25\textwidth}
    \includegraphics[width=\linewidth]{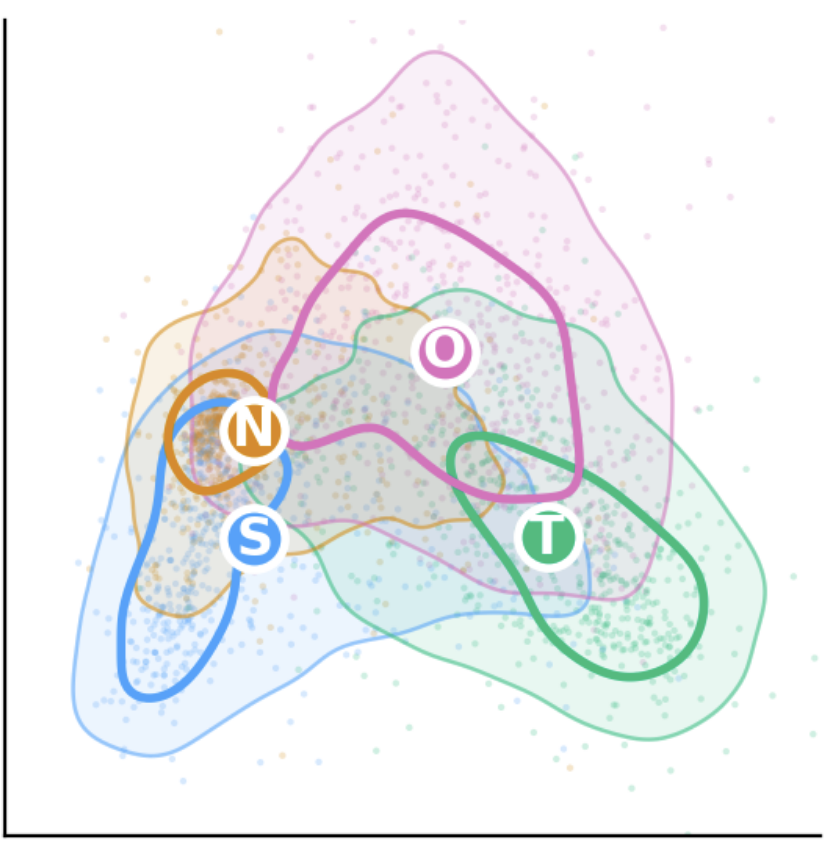}
    \caption{Per-speaker}\label{fig:emb-dicow}
  \end{subfigure}
  \hfil
  \begin{subfigure}{0.25\textwidth}
    \includegraphics[width=\linewidth]{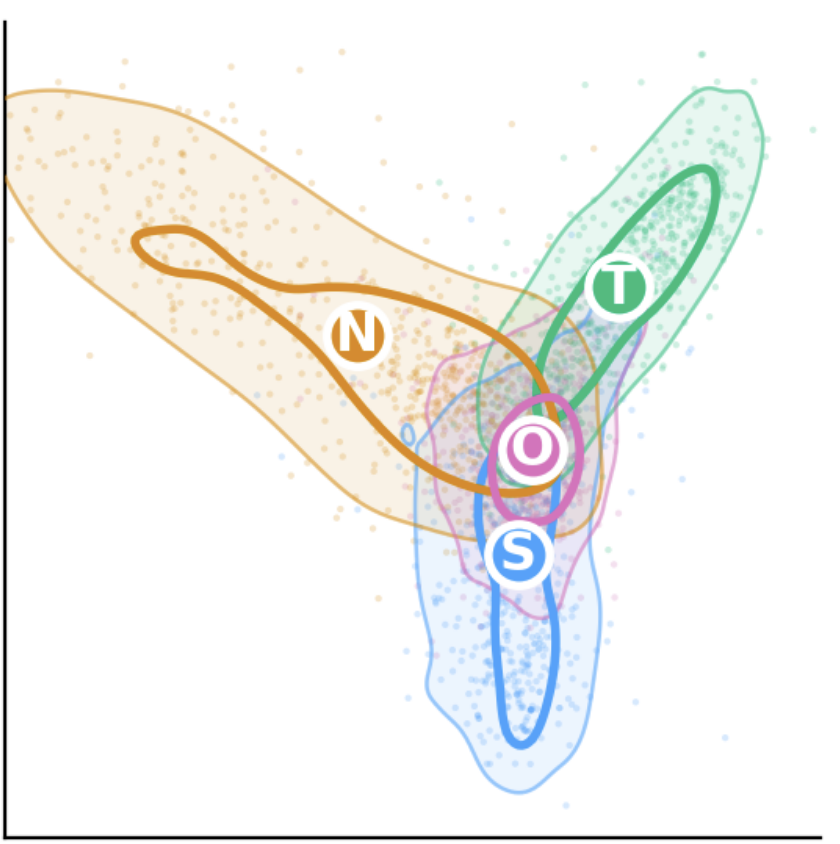}
    \caption{Standard heuristic \cite{lu2021-surt, sklyar2021-dat}}\label{fig:emb-std}
  \end{subfigure}
  \hfil
  \begin{subfigure}{0.25\textwidth}
    \includegraphics[width=\linewidth]{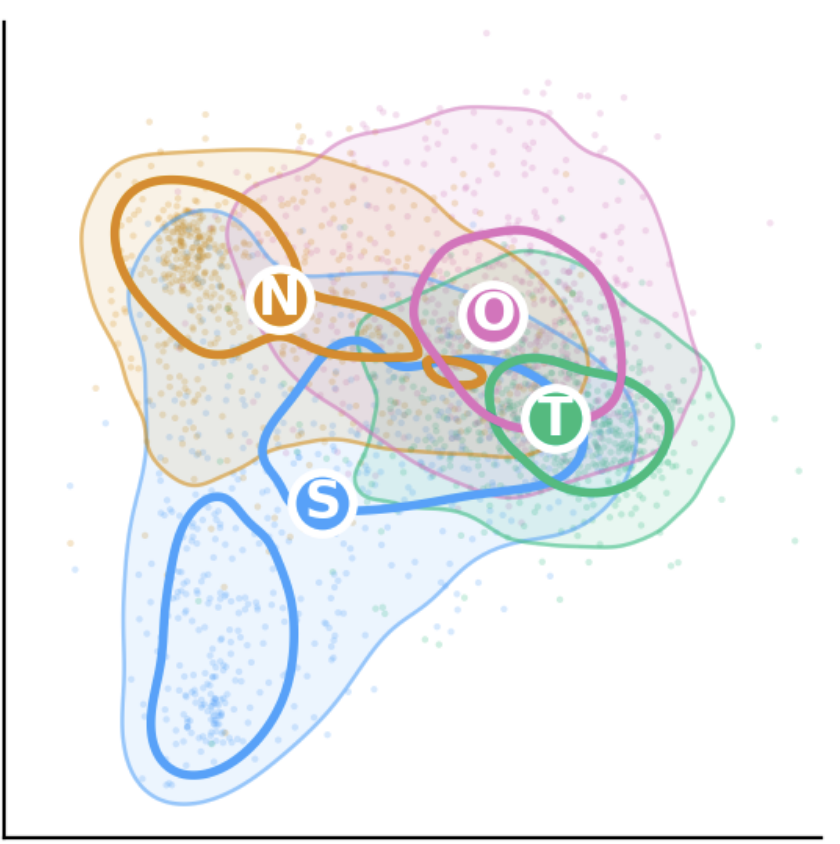}
    \caption{Speaker-continuity heuristic}\label{fig:emb-sc}
  \end{subfigure}
\caption{Final-layer encoder embeddings under oracle conditioning, projected onto the two leading Fisher discriminant directions of the final-layer Whisper encoder output: (a) per-speaker conditioning, the D-Orc configuration; (b) two-stream conditioning with the standard heuristic; and (c) two-stream conditioning with the heuristic of Section~\ref{sec:heuristic}, the H-Orc configuration of Table~\ref{tab:main}. Panel (b) is reported only here. Contours enclose 50\% and 90\% of each class over a subsample of the underlying bins.}
\label{fig:embedding-geometry}
\end{figure*}

\subsection{Predicting the HEAT Conditioning Mask}
\label{sec:results-mask}

Table~\ref{tab:frontend} reports frontend prediction quality for the mask predictor of Fig.~\ref{fig:frontend}(b) and for the diarization system whose output the cascaded construction consumes. DiariZen is shown twice: at the 16\,s window it was released with, and at the 30\,s window our mask predictor uses, since the mask has to span one ASR input. All terms are measured within a chunk, so they report local errors and do not indicate whether identities are linked correctly across the recording.

False alarm and missed speech compare directly across the two systems, since the union of the two streams is the total predicted speech. At the matched 30\,s window the predictor has slightly more false alarm than DiariZen, 4.20 against 3.75, and less missed speech, 4.57 against 5.87, despite observing each frame once rather than under ten overlapping windows. 

Confusion is not comparable across the two systems, since stream identity follows the assignment history and one wrong decision is charged for every frame after it. Within the predictor, building the objective up from PIL-CE lowers confusion from 10.91 to 7.50, and the swap counts show where that gain comes from. Unnecessary swaps decrease while missed swaps rise slightly, so the added objectives make the predictor hold its stream through an utterance rather than reproduce more of the heuristic's stream changes. False alarm and missed speech together also fall, so the stream objectives do not trade against detection.

\subsection{Conditioning Geometry in the Encoder}
\label{sec:results-conditioning}

Conditioning on a stream rather than on a speaker changes what the mask means without changing the interface it passes through, so it is not obvious that the recognizer represents the two alike. We examine this for the heuristic of Section~\ref{sec:heuristic}, whose effect on the recognizer was not measured in \cite{he2026scaling}. Holding content fixed and varying only the conditioning, we encode every one-second bin once per conditioning pass and take the mean-pooled final-layer Whisper encoder output. Figure~\ref{fig:embedding-geometry} projects these embeddings onto their two leading Fisher discriminant directions over eight meetings of forty 30-second chunks, with one contour per mask state. All three panels are given reference activity, so they differ only in how the mask is built: per-speaker in Fig.~\ref{fig:embedding-geometry}(a), and two assignment rules in (b) and (c).

Figure~\ref{fig:embedding-geometry}(a) is the reference. The four states are not reducible to speech presence: a mask recording only that speech is present would merge target with overlap and silence with non-target, and the recognizer would lose the distinction it is conditioned on. Under per-speaker conditioning the non-target state is separable from the states it would be merged with, though at much lower amplitude than the activity axis that dominates the projection.

Figure~\ref{fig:embedding-geometry}(c) shows that two-stream conditioning reproduces that structure. Substituting a target stream for a target speaker leaves the four states as distinguishable in the representation as per-speaker conditioning does, so the recognizer treats a stream target much as it treats a speaker target.

That equivalence is not automatic, and Fig.~\ref{fig:embedding-geometry}(b) shows it failing. Under the standard heuristic the non-target state overlaps heavily with both overlap and target, the four contours collapse toward a common region, and the mask reaches the encoder in effectively one state for most of its frames, which is close to no conditioning at all. Two passes conditioned on almost the same mask produce almost the same transcript, which is the collapse reported in \cite{he2026scaling}. The projection places that failure in the conditioning signal itself, before the decoder sees anything, though it does not establish why the conditioning parameters fail to differentiate the states during training.

Finally, we note that the standard heuristic attains higher separability than the rule of Fig.~\ref{fig:embedding-geometry}(c) along a purpose-fitted direction, despite collapsing the recognizer, so separability along a fitted direction should not be read as a measure of conditioning quality.

\begin{table*}[t]
\centering
\caption{\textbf{Recognition accuracy and inference speed}, beam-5 decoding. WER is tcORC-WER and SparseLM is SparseLibriMix. D and H denote per-speaker and HEAT conditioning of the DiCoW v3.3 recognizer of Section~\ref{implementation-details}, and Orc, Casc and Dir the three mask sources of Section~\ref{sec:configs}. AMI and ICSI are the SDM test splits, NOTSOFAR-1 the dev split, and CHiME-6 the eval split.}
\label{tab:main}
\small
\setlength{\tabcolsep}{8.0pt}
\begin{tabular}{lrrrrrrrrrrrr}
\toprule
& \multicolumn{2}{c}{AMI} & \multicolumn{2}{c}{ICSI} & \multicolumn{2}{c}{NOTSOFAR-1} & \multicolumn{2}{c}{CHiME-6} & \multicolumn{2}{c}{SparseLM $K{=}2$} & \multicolumn{2}{c}{SparseLM $K{=}3$} \\
\cmidrule(lr){2-3} \cmidrule(lr){4-5} \cmidrule(lr){6-7} \cmidrule(lr){8-9} \cmidrule(lr){10-11} \cmidrule(lr){12-13}
System & WER & RTFx & WER & RTFx & WER & RTFx & WER & RTFx & WER & RTFx & WER & RTFx \\
\midrule
\multicolumn{13}{l}{\emph{Oracle activity conditioning, RTFx on ASR}} \\
D-Orc  & 0.1663 & 16.7 & 0.2274 &  9.3 & 0.2607 &  9.1 & 0.5366 & 17.4 & 0.0393 & 4.6 & 0.1255 & 3.0 \\
H-Orc  & 0.1753 & 26.6 & 0.2397 & 23.7 & 0.2688 & 19.9 & 0.5252 & 19.8 & 0.0371 & 4.6 & 0.1736 & 4.1 \\
\midrule
\multicolumn{13}{l}{\emph{Predicted activity conditioning, RTFx on Frontend\,+\,ASR}} \\
D-Casc & 0.1740 &  8.7 & 0.2550 &  6.3 & 0.2555 &  6.3 & 0.4804 &  9.2 & 0.0449 & 3.6 & 0.2219 & 3.0 \\
H-Casc & 0.1769 & 12.5 & 0.2570 & 11.8 & 0.2566 & 11.0 & 0.4828 & 11.8 & 0.0428 & 3.3 & 0.2068 & 3.1 \\
\addlinespace
H-Dir  & 0.1911 & 15.6 & 0.3088 & 14.6 & 0.2760 & 12.6 & 0.5007 & 15.0 & 0.0567 & 3.7 & 0.2020 & 3.3 \\
\bottomrule
\end{tabular}
\end{table*}

\subsection{HEAT Speedup and Accuracy Cost}
\label{sec:results-main}

Table~\ref{tab:main} reports word error and inference speed for all five configurations, decoding with a beam of five throughout. The oracle rows recover the tradeoff reported in \cite{he2026scaling}: H-Orc runs 1.6 times faster than D-Orc on AMI and 2.6 times faster on ICSI, at a cost of 5.4\% relative word error on both. The speedup grows with the speaker inventory, since HEAT fixes the recognizer at two passes while per-speaker conditioning runs once per participant, and ICSI averages seven speakers per session against AMI's four. The accuracy cost is the price of a less definite target: a HEAT stream's overlap frames say that both streams are active, but not which the current pass should transcribe.

In the deployable comparison, H-Casc still runs 1.4 and 1.9 times faster than D-Casc, and the accuracy cost falls to 1.7\% relative on AMI and 0.8\% on ICSI. A missed or misattributed speaker costs per-speaker conditioning an entire recognition pass, whereas HEAT loses only the frames routed to the wrong stream. The two systems therefore converge as the frontend degrades rather than as it improves, so the gap measured under reference activity overstates what a deployed system pays.

\begin{figure}[t]
  \centering
  \includegraphics[width=\columnwidth]{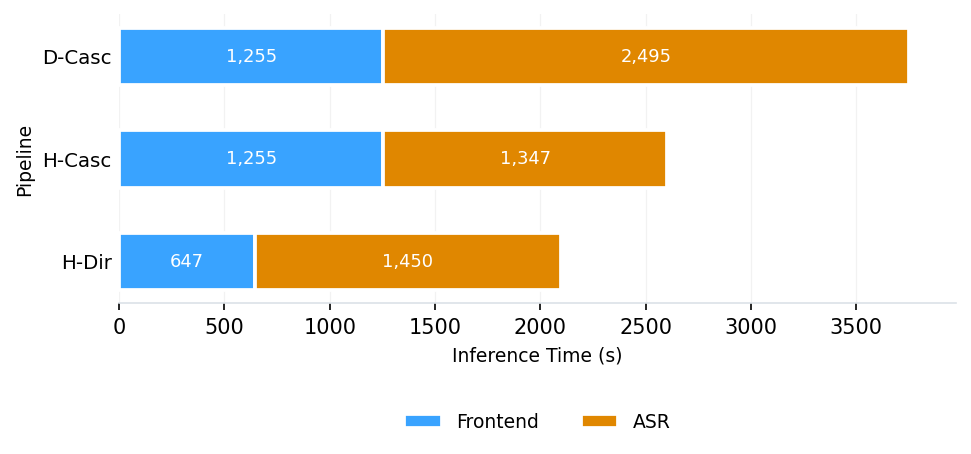}
\caption{Inference time in seconds on AMI-SDM test for D-Casc, H-Casc and H-Dir, with beam-5 decoding. The totals correspond to the AMI RTFx in Table~\ref{tab:main}, which gives inference speed on the remaining corpora.}
\label{fig:timing}
\end{figure}

The remaining corpora probe where the two-stream assumption holds. CHiME-6 has the highest rate of three-or-more-speaker overlap of any corpus we use, and H-Orc is ahead of D-Orc there, since dense overlap is where a per-speaker target is least well defined. Both cascaded configurations improve on their oracle counterparts on that corpus, which follows from the frontend coverage noted in Section~\ref{sec:data} rather than from anything the conditioning does.

SparseLibriMix controls the speaker count directly: H-Orc is ahead of D-Orc at $K{=}2$, where its mask coincides with the per-speaker mask, and behind at $K{=}3$, the smallest count at which speakers must share a stream, as also found in \cite{he2026scaling}. That inversion reverses between H-Casc and D-Casc, since the diarizer misses a speaker in 321 of 500 cuts at $K{=}3$. Per-speaker conditioning loses that speaker's pass, whereas HEAT, which never gave the speaker a stream of its own, loses less.

H-Dir removes the diarization pass, which buys a further 1.2 times speedup over H-Casc on AMI at a cost of 8.0\% relative word error, and leaves the pipeline with no diarizer at all. It is therefore the appropriate choice where a modest accuracy cost is acceptable, and it is the only configuration that removes the sliding window and the clustering pass, neither of which can run causally. Its limit is speaker count: on ICSI, which averages seven speakers per session against AMI's four, the cost rises to 20.2\%. H-Casc is within 1\% of D-Casc on that corpus, so this is a limit of the predictor rather than of two-stream conditioning, and the predictor is trained on corpora in the four-to-five range while the cascaded frontend resolves any number of speakers through clustering.

Figure~\ref{fig:timing} decomposes that speed into stages on AMI, showing that conditioning the recognizer on HEAT streams addresses only half of the problem. Moving from D-Casc to H-Casc cuts the ASR side almost in half but leaves the diarization pass untouched, so the frontend rises from a third of the chain to nearly half of it. H-Dir then halves the frontend, running the mask predictor once per chunk within the ASR pass rather than a sliding-window diarization followed by clustering, and brings the complete chain to 56\% of D-Casc.

\begin{table}[t]
\centering
\caption{\textbf{Speaker attribution under three conditioning sources.} Comparison of stream-agnostic and stream-aware orchestration on AMI-SDM test with greedy decoding, with each decode labeled by its own source-consistent diarization. H-Dir links no speakers to streams, so it has no stream-aware result. D-Orc and D-Casc obtain labels from their per-speaker passes and have no matching step.}
\label{tab:attribution}
\small
\begin{tabular}{lrrr}
\toprule
& & \multicolumn{2}{c}{tcpWER} \\
\cmidrule(lr){3-4}
Configuration & tcORC$\downarrow$ & agnostic$\downarrow$ & aware$\downarrow$ \\
\midrule
\multicolumn{4}{l}{\emph{Orchestration}} \\
H-Orc  & 0.1783 & 0.2017 & 0.1954 \\
H-Casc & 0.1806 & 0.2266 & 0.2184 \\
H-Dir  & 0.1901 & 0.2542 & --- \\
\midrule
\multicolumn{4}{l}{\emph{Built-in attribution}} \\
D-Orc  & 0.1673 & \multicolumn{2}{c}{0.1715} \\
D-Casc & 0.1754 & \multicolumn{2}{c}{0.2026} \\
\bottomrule
\end{tabular}
\end{table}

\subsection{Speaker Attribution}
\label{sec:results-attribution}

Table~\ref{tab:attribution} reports attributed word error under each conditioning source, with and without the stream restriction, alongside DiCoW's built-in attribution.

Interval matching fails where two speakers are co-active, and this is exactly the case HEAT is built to separate. The stream restriction helps for both H-Orc and H-Casc, with the larger gain for H-Casc, where predicted intervals produce more same-span ties than reference intervals do. The restriction removes the tied rival because the assignment placed it on the other stream, so the two-stream formulation that discards speaker identity is also what makes the identity recoverable at overlap.

The question is what that recovery costs relative to the speedup it comes with. Reading tcpWER against tcORC on the same output isolates the attribution penalty: D-Orc and D-Casc pay 0.0042 and 0.0272, against 0.0171 and 0.0378 for H-Orc and H-Casc with stream-aware orchestration. The per-speaker configurations are better on both, which is expected, since their labels come free from passes they were already running while HEAT recovers them from a pipeline trained for nothing else. The relevant comparison is that an orchestration step adding under 1\% of runtime closes most of the gap on a system running a fraction of the passes.

\section{Conclusion}
\label{sec:conclusion}

In this study, we showed that HEAT reduces inference cost on both halves of a SAC-ASR pipeline, in two steps. Conditioning DiCoW on two speaker-agnostic streams fixes the recognizer at two passes regardless of the number of participants, and with the same diarization frontend, H-Casc runs 1.4 and 1.9 times faster than D-Casc on AMI and ICSI for under 2\% relative word error. Comparing encoder representations across conditioning schemes, we found that the speaker-continuity heuristic keeps the four mask states as distinguishable as per-speaker conditioning does, whereas under the standard heuristic they merge.

That first step leaves the diarization pass untouched, and with it nearly half of the runtime. Because the two-stream mask carries no persistent speaker identity, a model built on a diarization backbone can instead predict it directly from the audio, emitting the masks of both streams in one pass. Permutation-invariant training alone lets that model collapse to VAD, which an added arrival-order objective prevents. The resulting H-Dir pipeline needs no sliding-window inference or speaker clustering, and cuts the AMI runtime to 56\% of D-Casc's at a 10\% relative word error cost, rising to 21\% on ICSI, the corpus with the most speakers per session.

We also explored recovering speaker attribution by matching decoded utterances to diarized speakers within each stream, which adds under 1\% of runtime and separates co-active speakers that timing alone cannot. Of the two deployable HEAT pipelines, attribution is better served by H-Casc, since global speaker clustering benefits from the sliding-window inference that direct prediction removes. H-Dir is therefore the appropriate choice when speaker-attributed output is not required, and H-Casc when it is.

Future work includes testing other backbones for the mask predictor and the recognizer, streaming ones in particular, where dropping long-term speaker tracking is most valuable but the heuristic must assign each utterance before the recording is complete. Additionally, the pipeline is amenable to end-to-end training, in which the stream assignment could be learned rather than reproduced from a fixed rule, and to combined decoding of the two streams. Serialized output training is a natural route to restoring speaker labels, letting the language model contribute to speaker prediction in place of interval matching. Finally, the two-speaker overlap assumption should be tested on other languages and conversational styles, and other heuristics may better capture conversational dynamics or bring the streams closer to the single-speaker distribution the backbone was pretrained on.

\bibliographystyle{IEEEtran}
\bibliography{ref}

@inproceedings{vonneumann2023-orcwer,
  author    = {von Neumann, Thilo and Boeddeker, Christoph and Kinoshita, Keisuke and Delcroix, Marc and Haeb-Umbach, Reinhold},
  title     = {On Word Error Rate Definitions and Their Efficient Computation for Multi-Speaker Speech Recognition Systems},
  booktitle = {IEEE International Conference on Acoustics, Speech and Signal Processing (ICASSP 2023)},
  pages     = {1--5},
  year      = {2023},
  doi       = {10.1109/ICASSP49357.2023.10094784}
}

@inproceedings{MeetEval23,
  author    = {von Neumann, Thilo and Boeddeker, Christoph and Delcroix, Marc and Haeb-Umbach, Reinhold},
  title     = {{MeetEval}: A Toolkit for Computation of Word Error Rates for Meeting Transcription Systems},
  booktitle = {7th International Workshop on Speech Processing in Everyday Environments (CHiME 2023)},
  pages     = {27--32},
  year      = {2023},
  doi       = {10.21437/CHiME.2023-6}
}

@inproceedings{he2026scaling,
  author    = {He, Xiluo and Polok, Alexander and Villalba, Jes{\'u}s and Thebaud, Thomas and Maciejewski, Matthew},
  title     = {Scaling multi-talker {ASR} with speaker-agnostic activity streams},
  booktitle = {IEEE International Conference on Acoustics, Speech and Signal Processing (ICASSP 2026)},
  pages     = {16727--16731},
  year      = {2026}
}

@article{polok2024-dicowdiarizationconditionedwhispertarget,
  author    = {Alexander Polok and Dominik Klement and Martin Kocour and Jiangyu Han and Federico Landini and Bolaji Yusuf and Matthew Wiesner and Sanjeev Khudanpur and Jan Černocký and Lukáš Burget},
  title     = {{DiCoW}: Diarization-conditioned {Whisper} for target speaker automatic speech recognition},
  journal   = {Computer Speech \& Language},
  volume    = {95},
  pages     = {101841},
  year      = {2026},
  doi       = {10.1016/j.csl.2025.101841}
}

@inproceedings{polok2024-targetspeakerasrwhisper,
  author    = {Polok, Alexander and Klement, Dominik and Wiesner, Matthew and Khudanpur, Sanjeev and Černocký, Jan and Burget, Lukáš},
  title     = {Target Speaker {ASR} with {Whisper}},
  booktitle = {IEEE International Conference on Acoustics, Speech and Signal Processing (ICASSP 2025)},
  pages     = {1--5},
  year      = {2025},
  doi       = {10.1109/ICASSP49660.2025.10887683}
}

@inproceedings{wang25-ssa,
  author    = {Weiqing Wang and Taejin Park and Ivan Medennikov and Jinhan Wang and Kunal Dhawan and He Huang and Koluguri, Nithin Rao and Jagadeesh Balam and Boris Ginsburg},
  title     = {Speaker Targeting via Self-Speaker Adaptation for Multi-talker {ASR}},
  booktitle = {Interspeech 2025},
  pages     = {5498--5502},
  year      = {2025},
  doi       = {10.21437/Interspeech.2025-2142}
}

@inproceedings{medennikov20-chime,
  author    = {Ivan Medennikov and Maxim Korenevsky and Tatiana Prisyach and Yuri Khokhlov and Mariya Korenevskaya and Ivan Sorokin and Tatiana Timofeeva and Anton Mitrofanov and Andrei Andrusenko and Ivan Podluzhny and Aleksandr Laptev and Aleksei Romanenko},
  title     = {The {STC} System for the {CHiME-6} Challenge},
  booktitle = {6th International Workshop on Speech Processing in Everyday Environments (CHiME 2020)},
  pages     = {36--41},
  year      = {2020},
  doi       = {10.21437/CHiME.2020-9}
}

@inproceedings{sklyar2021-dat,
  author    = {Sklyar, Ilya and Piunova, Anna and Liu, Yulan},
  title     = {Streaming multi-speaker {ASR} with {RNN-T}},
  booktitle = {IEEE International Conference on Acoustics, Speech and Signal Processing (ICASSP 2021)},
  pages     = {6903--6907},
  year      = {2021}
}

@article{raj2023-surt2,
  author    = {Raj, Desh and Povey, Daniel and Khudanpur, Sanjeev},
  title     = {{SURT} 2.0: Advances in transducer-based multi-talker speech recognition},
  journal   = {IEEE/ACM Transactions on Audio, Speech, and Language Processing},
  volume    = {31},
  pages     = {3800--3813},
  year      = {2023}
}

@inproceedings{kalda2024pixit,
  author    = {Joonas Kalda and Clément Pagés and Ricard Marxer and Tanel Alumäe and Hervé Bredin},
  title     = {{PixIT}: Joint Training of Speaker Diarization and Speech Separation from Real-world Multi-speaker Recordings},
  booktitle = {The Speaker and Language Recognition Workshop (Odyssey 2024)},
  pages     = {115--122},
  year      = {2024},
  doi       = {10.21437/odyssey.2024-17}
}

@inproceedings{chen2023speech,
  author    = {Chen, Zhuo and Kanda, Naoyuki and Wu, Jian and Wu, Yu and Wang, Xiaofei and Yoshioka, Takuya and Li, Jinyu and Sivasankaran, Sunit and Eskimez, Sefik Emre},
  title     = {Speech separation with large-scale self-supervised learning},
  booktitle = {IEEE International Conference on Acoustics, Speech and Signal Processing (ICASSP 2023)},
  pages     = {1--5},
  year      = {2023}
}

@inproceedings{yoshioka2018recognizing,
  author    = {Takuya Yoshioka and Hakan Erdogan and Zhuo Chen and Xiong Xiao and Fil Alleva},
  title     = {Recognizing Overlapped Speech in Meetings: A Multichannel Separation Approach Using Neural Networks},
  booktitle = {Interspeech 2018},
  pages     = {3038--3042},
  year      = {2018},
  doi       = {10.21437/Interspeech.2018-2284}
}

@inproceedings{saon2013speaker,
  author    = {Saon, George and Soltau, Hagen and Nahamoo, David and Picheny, Michael},
  title     = {Speaker adaptation of neural network acoustic models using i-vectors},
  booktitle = {IEEE Automatic Speech Recognition and Understanding Workshop (ASRU 2013)},
  pages     = {55--59},
  year      = {2013}
}

@inproceedings{kanda2020serialized,
  author    = {Naoyuki Kanda and Yashesh Gaur and Xiaofei Wang and Zhong Meng and Takuya Yoshioka},
  title     = {Serialized Output Training for End-to-End Overlapped Speech Recognition},
  booktitle = {Interspeech 2020},
  pages     = {2797--2801},
  year      = {2020},
  doi       = {10.21437/Interspeech.2020-999}
}

@inproceedings{raj2021integration,
  author    = {Raj, Desh and Denisov, Pavel and Chen, Zhuo and Erdogan, Hakan and Huang, Zili and He, Maokui and Watanabe, Shinji and Du, Jun and Yoshioka, Takuya and Luo, Yi and others},
  title     = {Integration of speech separation, diarization, and recognition for multi-speaker meetings: System description, comparison, and analysis},
  booktitle = {IEEE Spoken Language Technology Workshop (SLT 2021)},
  pages     = {897--904},
  year      = {2021}
}

@inproceedings{kanda2019simutaneous,
  author    = {Kanda, Naoyuki and Horiguchi, Shota and Fujita, Yusuke and Xue, Yawen and Nagamatsu, Kenji and Watanabe, Shinji},
  title     = {Simultaneous Speech Recognition and Speaker Diarization for Monaural Dialogue Recordings with Target-Speaker Acoustic Models},
  booktitle = {IEEE Automatic Speech Recognition and Understanding Workshop (ASRU 2019)},
  pages     = {31--38},
  year      = {2019},
  doi       = {10.1109/ASRU46091.2019.9004009}
}

@inproceedings{radford2023robust,
  author    = {Radford, Alec and Kim, Jong Wook and Xu, Tao and Brockman, Greg and McLeavey, Christine and Sutskever, Ilya},
  title     = {Robust speech recognition via large-scale weak supervision},
  booktitle = {International Conference on Machine Learning (ICML 2023)},
  pages     = {28492--28518},
  year      = {2023}
}

@article{prabhavalkar2023end,
  author    = {Prabhavalkar, Rohit and Hori, Takaaki and Sainath, Tara N and Schl{\"u}ter, Ralf and Watanabe, Shinji},
  title     = {End-to-end speech recognition: A survey},
  journal   = {IEEE/ACM Transactions on Audio, Speech, and Language Processing},
  volume    = {32},
  pages     = {325--351},
  year      = {2023}
}

@article{li2022recent,
  author    = {Li, Jinyu and others},
  title     = {Recent advances in end-to-end automatic speech recognition},
  journal   = {APSIPA Transactions on Signal and Information Processing},
  volume    = {11},
  number    = {1},
  year      = {2022}
}

@inproceedings{meng2024empowering,
  author    = {Lingwei Meng and Jiawen Kang and Yuejiao Wang and Zengrui Jin and Xixin Wu and Xunying Liu and Helen Meng},
  title     = {Empowering {Whisper} as a Joint Multi-Talker and Target-Talker Speech Recognition System},
  booktitle = {Interspeech 2024},
  pages     = {4653--4657},
  year      = {2024},
  doi       = {10.21437/Interspeech.2024-971}
}

@inproceedings{cornell2024chime,
  author    = {Samuele Cornell and Tae Jin Park and He Huang and Christoph Boeddeker and Xuankai Chang and Matthew Maciejewski and Matthew S Wiesner and Paola Garcia and Shinji Watanabe},
  title     = {The {CHiME-8} {DASR} Challenge for Generalizable and Array Agnostic Distant Automatic Speech Recognition and Diarization},
  booktitle = {8th International Workshop on Speech Processing in Everyday Environments (CHiME 2024)},
  pages     = {1--6},
  year      = {2024},
  doi       = {10.21437/CHiME.2024-1}
}

@inproceedings{yu2017recognizing,
  author    = {Dong Yu and Xuankai Chang and Yanmin Qian},
  title     = {Recognizing Multi-Talker Speech with Permutation Invariant Training},
  booktitle = {Interspeech 2017},
  pages     = {2456--2460},
  year      = {2017},
  doi       = {10.21437/Interspeech.2017-305}
}

@inproceedings{yu2017permutation,
  author    = {Yu, Dong and Kolb{\ae}k, Morten and Tan, Zheng-Hua and Jensen, Jesper},
  title     = {Permutation invariant training of deep models for speaker-independent multi-talker speech separation},
  booktitle = {IEEE International Conference on Acoustics, Speech and Signal Processing (ICASSP 2017)},
  pages     = {241--245},
  year      = {2017}
}

@article{lu2021-surt,
  author    = {Lu, Liang and Kanda, Naoyuki and Li, Jinyu and Gong, Yifan},
  title     = {Streaming end-to-end multi-talker speech recognition},
  journal   = {IEEE Signal Processing Letters},
  volume    = {28},
  pages     = {803--807},
  year      = {2021}
}

@article{abramovski2025-chime9nsfsummary,
  author    = {Abramovski, Igor and Vinnikov, Alon and Shaer, Shalev and Kanda, Naoyuki and Wang, Xiaofei and Ivry, Amir and Krupka, Eyal},
  title     = {Summary of the {NOTSOFAR-1} challenge: Highlights and learnings},
  journal   = {Computer Speech \& Language},
  volume    = {93},
  pages     = {101796},
  year      = {2025}
}

@inproceedings{han2025diarizen,
  author    = {Han, Jiangyu and Landini, Federico and Rohdin, Johan and Silnova, Anna and Diez, Mireia and Burget, Lukáš},
  title     = {Leveraging Self-Supervised Learning for Speaker Diarization},
  booktitle = {IEEE International Conference on Acoustics, Speech and Signal Processing (ICASSP 2025)},
  pages     = {1--5},
  year      = {2025},
  doi       = {10.1109/ICASSP49660.2025.10889475}
}

@inproceedings{kraaij2005ami,
  author    = {Kraaij, Wessel and Hain, Thomas and Lincoln, Mike and Post, Wilfried},
  title     = {The {AMI} meeting corpus},
  booktitle = {International Conference on Methods and Techniques in Behavioral Research (Measuring Behavior 2005)},
  pages     = {1--4},
  year      = {2005}
}

@inproceedings{janin2003icsi,
  author    = {Janin, Adam and Baron, Don and Edwards, Jane and Ellis, Dan and Gelbart, David and Morgan, Nelson and Peskin, Barbara and Pfau, Thilo and Shriberg, Elizabeth and Stolcke, Andreas and others},
  title     = {The {ICSI} meeting corpus},
  booktitle = {IEEE International Conference on Acoustics, Speech and Signal Processing (ICASSP 2003)},
  pages     = {I{\char45}364--I{\char45}367},
  year      = {2003}
}

@misc{cosentino2020librimix,
  author       = {Cosentino, Joris and Pariente, Manuel and Cornell, Samuele and Deleforge, Antoine and Vincent, Emmanuel},
  title        = {{LibriMix}: An open-source dataset for generalizable speech separation},
  howpublished = {arXiv:2005.11262},
  year         = {2020}
}

@inproceedings{subramanian2025improving,
  author    = {Aswin Shanmugam Subramanian and Amit Das and Naoyuki Kanda and Jinyu Li and Xiaofei Wang and Yifan Gong},
  title     = {Improving Practical Aspects of End-to-End Multi-Talker Speech Recognition for Online and Offline Scenarios},
  booktitle = {Interspeech 2025},
  pages     = {5508--5512},
  year      = {2025},
  doi       = {10.21437/Interspeech.2025-2414}
}

@inproceedings{kanda2022-streamingembedding,
  author    = {Naoyuki Kanda and Jian Wu and Yu Wu and Xiong Xiao and Zhong Meng and Xiaofei Wang and Yashesh Gaur and Zhuo Chen and Jinyu Li and Takuya Yoshioka},
  title     = {Streaming Speaker-Attributed {ASR} with Token-Level Speaker Embeddings},
  booktitle = {Interspeech 2022},
  pages     = {521--525},
  year      = {2022},
  doi       = {10.21437/Interspeech.2022-253}
}

@inproceedings{linke2025context,
  author    = {Linke, Julian and Winkler, Jana and Schuppler, Barbara},
  title     = {Context is all you need? {L}ow-resource conversational {ASR} profits from context, coming from the same or from the other speaker},
  booktitle = {Interspeech 2025},
  year      = {2025}
}

@inproceedings{meng2023unified,
  author    = {Lingwei Meng and Jiawen Kang and Mingyu Cui and Haibin Wu and Xixin Wu and Helen Meng},
  title     = {Unified Modeling of Multi-Talker Overlapped Speech Recognition and Diarization with a Sidecar Separator},
  booktitle = {Interspeech 2023},
  pages     = {3467--3471},
  year      = {2023},
  doi       = {10.21437/Interspeech.2023-1422}
}

@inproceedings{masumura2023end,
  author    = {Ryo Masumura and Naoki Makishima and Taiga Yamane and Yoshihiko Yamazaki and Saki Mizuno and Mana Ihori and Mihiro Uchida and Keita Suzuki and Hiroshi Sato and Tomohiro Tanaka and Akihiko Takashima and Satoshi Suzuki and Takafumi Moriya and Nobukatsu Hojo and Atsushi Ando},
  title     = {End-to-End Joint Target and Non-Target Speakers {ASR}},
  booktitle = {Interspeech 2023},
  pages     = {2903--2907},
  year      = {2023},
  doi       = {10.21437/Interspeech.2023-1640}
}

@inproceedings{cetin06_interspeech,
  author    = {{\"O}zg{\"u}r {\c{C}}etin and Elizabeth Shriberg},
  title     = {Analysis of overlaps in meetings by dialog factors, hot spots, speakers, and collection site: Insights for automatic speech recognition},
  booktitle = {Interspeech 2006},
  year      = {2006},
  doi       = {10.21437/Interspeech.2006-91}
}

@inproceedings{kanda22_interspeech,
  author    = {Naoyuki Kanda and Jian Wu and Yu Wu and Xiong Xiao and Zhong Meng and Xiaofei Wang and Yashesh Gaur and Zhuo Chen and Jinyu Li and Takuya Yoshioka},
  title     = {Streaming Multi-Talker {ASR} with Token-Level Serialized Output Training},
  booktitle = {Interspeech 2022},
  pages     = {3774--3778},
  year      = {2022},
  doi       = {10.21437/Interspeech.2022-7}
}

@article{vonneumann25_tcorcwer,
  author    = {von Neumann, Thilo and Boeddeker, Christoph and Delcroix, Marc and Haeb-Umbach, Reinhold},
  title     = {Word Error Rate Definitions and Algorithms for Long-Form Multi-Talker Speech Recognition},
  journal   = {IEEE Transactions on Audio, Speech, and Language Processing},
  volume    = {33},
  pages     = {3174--3188},
  year      = {2025},
  doi       = {10.1109/TASLPRO.2025.3589862}
}

@inproceedings{polok2026_sedicow,
  author    = {Polok, Alexander and Klement, Dominik and Cornell, Samuele and Wiesner, Matthew and Černocký, Jan and Khudanpur, Sanjeev and Burget, Lukáš},
  title     = {{SE-DiCoW}: Self-Enrolled Diarization-Conditioned {Whisper}},
  booktitle = {IEEE International Conference on Acoustics, Speech and Signal Processing (ICASSP 2026)},
  year      = {2026}
}

@inproceedings{shafey2019joint,
  author    = {Laurent {El Shafey} and Hagen Soltau and Izhak Shafran},
  title     = {Joint Speech Recognition and Speaker Diarization via Sequence Transduction},
  booktitle = {Interspeech 2019},
  pages     = {396--400},
  year      = {2019},
  doi       = {10.21437/Interspeech.2019-1943}
}

@inproceedings{cornell2024joint,
  author    = {Cornell, Samuele and Jung, Jee-Weon and Watanabe, Shinji and Squartini, Stefano},
  title     = {One Model to Rule Them All? {T}owards End-to-End Joint Speaker Diarization and Speech Recognition},
  booktitle = {IEEE International Conference on Acoustics, Speech and Signal Processing (ICASSP 2024)},
  pages     = {11856--11860},
  year      = {2024},
  doi       = {10.1109/ICASSP48485.2024.10447957}
}

@misc{kocour2025adapting,
  author       = {Martin Kocour and Martin Karafiat and Alexander Polok and Dominik Klement and Lukáš Burget and Jan Černocký},
  title        = {Adapting Diarization-Conditioned {Whisper} for End-to-End Multi-Talker Speech Recognition},
  howpublished = {arXiv:2510.03723},
  year         = {2026}
}

@inproceedings{meng2025llm,
  author    = {Meng, Lingwei and Hu, Shujie and Kang, Jiawen and Li, Zhaoqing and Wang, Yuejiao and Wu, Wenxuan and Wu, Xixin and Liu, Xunying and Meng, Helen},
  title     = {Large Language Model Can Transcribe Speech in Multi-Talker Scenarios with Versatile Instructions},
  booktitle = {IEEE International Conference on Acoustics, Speech and Signal Processing (ICASSP 2025)},
  pages     = {1--5},
  year      = {2025},
  doi       = {10.1109/ICASSP49660.2025.10887642}
}

@inproceedings{shi2025serialized,
  author    = {Shi, Hao and Fujita, Yusuke and Mizumoto, Tomoya and Liu, Lianbo and Kojima, Atsushi and Sudo, Yui},
  title     = {Serialized Output Prompting for Large Language Model-based Multi-Talker Speech Recognition},
  booktitle = {IEEE Automatic Speech Recognition and Understanding Workshop (ASRU 2025)},
  pages     = {1--8},
  year      = {2025},
  doi       = {10.1109/ASRU65441.2025.11434801}
}

@misc{huo2026tag,
  author       = {Mingyue Huo and Yiwen Shao and Yuheng Zhang},
  title        = {{TagSpeech}: End-to-End Multi-Speaker {ASR} and Diarization with Fine-Grained Temporal Grounding},
  howpublished = {arXiv:2601.06896},
  year         = {2026}
}

@inproceedings{wang2024resource,
  author    = {Wang, Weiqing and Dhawan, Kunal and Park, Taejin and Puvvada, Krishna C. and Medennikov, Ivan and Majumdar, Somshubra and Huang, He and Balam, Jagadeesh and Ginsburg, Boris},
  title     = {Resource-Efficient Adaptation of Speech Foundation Models for Multi-Speaker {ASR}},
  booktitle = {IEEE Spoken Language Technology Workshop (SLT 2024)},
  pages     = {1224--1231},
  year      = {2024},
  doi       = {10.1109/SLT61566.2024.10832215}
}

@inproceedings{park2025sortformer,
  author    = {Park, Taejin and Medennikov, Ivan and Dhawan, Kunal and Wang, Weiqing and Huang, He and Koluguri, Nithin Rao and Puvvada, Krishna C and Balam, Jagadeesh and Ginsburg, Boris},
  title     = {{Sortformer}: A Novel Approach for Permutation-Resolved Speaker Supervision in Speech-to-Text Systems},
  booktitle = {International Conference on Machine Learning (ICML 2025)},
  pages     = {48153--48169},
  year      = {2025}
}

@inproceedings{raj24_odyssey,
  author    = {Desh Raj and Matthew Wiesner and Matthew Maciejewski and Paola Garcia and Daniel Povey and Sanjeev Khudanpur},
  title     = {On Speaker Attribution with {SURT}},
  booktitle = {The Speaker and Language Recognition Workshop (Odyssey 2024)},
  pages     = {91--98},
  year      = {2024},
  doi       = {10.21437/odyssey.2024-14}
}

@inproceedings{watanabe20b_chime,
  author    = {Shinji Watanabe and Michael Mandel and Jon Barker and Emmanuel Vincent and Ashish Arora and Xuankai Chang and Sanjeev Khudanpur and Vimal Manohar and Daniel Povey and Desh Raj and David Snyder and Aswin Shanmugam Subramanian and Jan Trmal and Bar Ben Yair and Christoph Boeddeker and Zhaoheng Ni and Yusuke Fujita and Shota Horiguchi and Naoyuki Kanda and Takuya Yoshioka and Neville Ryant},
  title     = {{CHiME-6} Challenge: Tackling Multispeaker Speech Recognition for Unsegmented Recordings},
  booktitle = {6th International Workshop on Speech Processing in Everyday Environments (CHiME 2020)},
  pages     = {1--7},
  year      = {2020},
  doi       = {10.21437/CHiME.2020-1}
}

@inproceedings{cornell23_chime,
  author    = {Samuele Cornell and Matthew S. Wiesner and Shinji Watanabe and Desh Raj and Xuankai Chang and Paola Garcia and Yoshiki Masuyama and Zhong-Qiu Wang and Stefano Squartini and Sanjeev Khudanpur},
  title     = {The {CHiME-7} {DASR} Challenge: Distant Meeting Transcription with Multiple Devices in Diverse Scenarios},
  booktitle = {7th International Workshop on Speech Processing in Everyday Environments (CHiME 2023)},
  pages     = {1--6},
  year      = {2023},
  doi       = {10.21437/CHiME.2023-1}
}

@inproceedings{wang2025_metacat,
  author    = {Wang, Jinhan and Wang, Weiqing and Dhawan, Kunal and Park, Taejin and Kim, Myungjong and Medennikov, Ivan and Huang, He and Koluguri, Nithin Rao and Balam, Jagadeesh and Ginsburg, Boris},
  title     = {{META-CAT}: Speaker-Informed Speech Embeddings via Meta Information Concatenation for Multi-talker {ASR}},
  booktitle = {IEEE International Conference on Acoustics, Speech and Signal Processing (ICASSP 2025)},
  pages     = {1--5},
  year      = {2025},
  doi       = {10.1109/ICASSP49660.2025.10889841}
}

@misc{snyder2015_musan,
  author       = {David Snyder and Guoguo Chen and Daniel Povey},
  title        = {{MUSAN}: A Music, Speech, and Noise Corpus},
  howpublished = {arXiv:1510.08484},
  year         = {2015}
}

@inproceedings{panayotov2015_librispeech,
  author    = {Panayotov, Vassil and Chen, Guoguo and Povey, Daniel and Khudanpur, Sanjeev},
  title     = {{Librispeech}: An {ASR} corpus based on public domain audio books},
  booktitle = {IEEE International Conference on Acoustics, Speech and Signal Processing (ICASSP 2015)},
  pages     = {5206--5210},
  year      = {2015},
  doi       = {10.1109/ICASSP.2015.7178964}
}

@inproceedings{wang2023_wespeaker,
  author    = {Wang, Hongji and Liang, Chengdong and Wang, Shuai and Chen, Zhengyang and Zhang, Binbin and Xiang, Xu and Deng, Yanlei and Qian, Yanmin},
  title     = {{Wespeaker}: A Research and Production Oriented Speaker Embedding Learning Toolkit},
  booktitle = {IEEE International Conference on Acoustics, Speech and Signal Processing (ICASSP 2023)},
  pages     = {1--5},
  year      = {2023},
  doi       = {10.1109/ICASSP49357.2023.10096626}
}

@inproceedings{wang2024_diarizationlm,
  author    = {Quan Wang and Yiling Huang and Guanlong Zhao and Evan Clark and Wei Xia and Hank Liao},
  title     = {{DiarizationLM}: Speaker Diarization Post-Processing with Large Language Models},
  booktitle = {Interspeech 2024},
  pages     = {3754--3758},
  year      = {2024},
  doi       = {10.21437/Interspeech.2024-209}
}

@inproceedings{Yu2022M2MeT,
  author    = {Yu, Fan and Zhang, Shiliang and Fu, Yihui and Xie, Lei and Zheng, Siqi and Du, Zhihao and Huang, Weilong and Guo, Pengcheng and Yan, Zhijie and Ma, Bin and Xu, Xin and Bu, Hui},
  title     = {{M2MeT}: The {ICASSP} 2022 Multi-Channel Multi-Party Meeting Transcription Challenge},
  booktitle = {IEEE International Conference on Acoustics, Speech and Signal Processing (ICASSP 2022)},
  year      = {2022}
}

@inproceedings{Yu2022Summary,
  author    = {Yu, Fan and Zhang, Shiliang and Guo, Pengcheng and Fu, Yihui and Du, Zhihao and Zheng, Siqi and Huang, Weilong and Xie, Lei and Tan, Zheng-Hua and Wang, DeLiang and Qian, Yanmin and Lee, Kong Aik and Yan, Zhijie and Ma, Bin and Xu, Xin and Bu, Hui},
  title     = {Summary on the {ICASSP} 2022 Multi-Channel Multi-Party Meeting Transcription Grand Challenge},
  booktitle = {IEEE International Conference on Acoustics, Speech and Signal Processing (ICASSP 2022)},
  year      = {2022}
}

@inproceedings{plaquet23_interspeech,
  author    = {Alexis Plaquet and Hervé Bredin},
  title     = {Powerset multi-class cross entropy loss for neural speaker diarization},
  booktitle = {Interspeech 2023},
  pages     = {3222--3226},
  year      = {2023},
  doi       = {10.21437/Interspeech.2023-205}
}

@inproceedings{fu21b_interspeech,
  author    = {Yihui Fu and Luyao Cheng and Shubo Lv and Yukai Jv and Yuxiang Kong and Zhuo Chen and Yanxin Hu and Lei Xie and Jian Wu and Hui Bu and Xin Xu and Jun Du and Jingdong Chen},
  title     = {{AISHELL-4}: An Open Source Dataset for Speech Enhancement, Separation, Recognition and Speaker Diarization in Conference Scenario},
  booktitle = {Interspeech 2021},
  pages     = {3665--3669},
  year      = {2021},
  doi       = {10.21437/Interspeech.2021-1397}
}

\end{document}